\documentstyle[epsf,seceq,amsbsy]{ptptex}
\notypesetlogo
\newcommand {\beq}{\begin{equation}}
\newcommand {\eeq}{\end{equation}}
\newcommand {\beqa}{\begin{eqnarray}}
\newcommand {\eeqa}{\end{eqnarray}}
\newcommand {\beqan}{\begin{eqnarray*}}
\newcommand {\eeqan}{\end{eqnarray*}}
\newcommand {\n}{\nonumber \\}

\def\ep{\epsilon}
\def\th{\theta}
\def\la{\lambda}
\def\da{\dot{a}}

 \title{ IIB Matrix Model\footnote{Based on the talk given by H. Kawai at the 13th Nishinomiya-Yukawa
Memorial Symposium ``Dynamics of Fields and Strings'' (November 12-13, 1998)
and on the talk by S. Iso at the YITP workshop (November 16-18, 1998)
        } }

 \author{          Hajime {\sc Aoki},$^{1}$$^,$\footnote
           {e-mail address : haoki@ccthmail.kek.jp}
           Satoshi {\sc Iso},$^{1}$$^,$\footnote
           {e-mail address : satoshi.iso@kek.jp}
           Hikaru {\sc Kawai},$^{2}$$^,$\footnote
           {e-mail address : hkawai@gauge.scphys.kyoto-u.ac.jp}
           Yoshihisa {\sc Kitazawa},$^{1}$$^,$\footnote
           {e-mail address : kitazawa@post.kek.jp}
           Asato {\sc Tsuchiya}$^{3}$$^,$\footnote
           {e-mail address : tsuchiya@funpth.phys.sci.osaka-u.ac.jp}
 and Tsukasa {\sc Tada}$^{4}$$^,$\footnote
           {e-mail address : tada@tohwa-u.ac.jp}}
\inst{         $^{1}$ High Energy Accelerator Research Organization,
             KEK\\ 
 Tsukuba 305-0801, Japan \\
        $^{2}$ Department of Physics, Kyoto University, Kyoto 606-8502, Japan\\

        $^{3}$ Department of Physics, Osaka University, Toyonaka 560-0043, Japan\\
        $^{4}$ Tohwa Institute for Science, Tohwa University, 
                  Fukuoka 815-8510, Japan}

\abst
{
We review our proposal for a constructive definition of superstring,
type IIB matrix model. The IIB matrix model is a manifestly covariant
model for space-time and matter which possesses ${\cal N}=2$ supersymmetry
in ten dimensions. We refine our arguments to reproduce string perturbation
theory based on the loop equations. We emphasize that the space-time is
dynamically determined from the eigenvalue distributions of the matrices.
We also explain how matter, gauge fields and gravitation appear as
fluctuations around dynamically determined space-time.
}

\begin{document}

\maketitle

\section{Introduction}
Several proposals have been made  as constructive definitions
of  superstring theory.
\cite{IKKT,BFSS,DVV,periwal,yoneya,Polch,sugawara,Itoyama,hetero,IIBvariant}
The type-IIB matrix model, \cite{IKKT,FKKT,AIKKT}
a large $N$ reduced model of maximally supersymmetric Yang-Mills theory,
is one of those proposals.
It is defined by the following action:
\beq
S  =  -{1\over g^2}{\rm Tr}\left({1\over 4}[A_{\mu},A_{\nu}][A^{\mu},A^{\nu}]
+{1\over 2}\bar{\psi}\Gamma ^{\mu}[A_{\mu},\psi ]\right) ,
\label{action}
\eeq
where
$A_{\mu}$ and $\psi$ are $N \times N$ Hermitian matrices,
the former is a ten-dimensional vector and  the latter is a ten-dimensional
Majorana-Weyl spinor field respectively.
It is formulated in a manifestly covariant way,  which
is suitable for studying nonperturbative issues of superstring theory.
Since it is a simple  model of matrices in zero dimension,
it does not possess degenerate vacua unlike its higher dimensional cousins.
It is possible that the model possesses a unique vacuum, namely our space-time.
If so, we can in principle predict the  dimensionality of
the space-time, low-energy gauge group and  matter contents by solving this
model.
In such an endeavor, this model can be studied by numerical simulations
effectively.
In this paper, we review the IIB matrix model and explain how
the space-time appears dynamically and how the low energy gauge
symmetry and diffeomorphism invariance emerge microscopically.
\par
We first list several important properties of the IIB matrix model.
This model can be  regarded as a large $N$ reduced model of
ten-dimensional ${\cal N}=1$
supersymmetric $SU(N)$ Yang-Mills theory.
It was shown\cite{RM} that a large $N$ gauge theory
can be equivalently described by its reduced  model, namely  a model
defined on a single point. In this reduction procedure,  a space-time
translation is represented in the color $SU(N)$ space, and
the eigenvalues
of the matrices are interpreted as the momenta of fields.
Therefore, the basic assumption in this identification is that the
eigenvalues are
uniformly distributed.
As a constructive definition of a superstring, on the other hand,
we will see  that we need to interpret the eigenvalues of matrices
as the coordinates of space-time points.  The  interpretation
is T-dual to the above.
\par
Since our IIB matrix model is defined on a single point,
the commutator of the  supersymmetry which we inherit from the
ten-dimensional ${\cal N}=1$
supersymmetric $SU(N)$ Yang-Mills theory,
\beq
\cases{
\delta^{(1)} A_\mu &  $= i \bar{\epsilon_1}\Gamma_\mu \psi $\cr
\delta^{(1)} \psi & $=\frac{i}{2}\Gamma^{\mu\nu} [A^\mu, A^\nu] {\epsilon_1}$,
}  \label{Ssym1}
\eeq
vanishes up to a field-dependent
gauge transformation, and we can no longer interpret
this supersymmetry as  space-time supersymmetry in the original sense.
However, after the reduction, we acquire  an extra bosonic symmetry,
\begin{equation}
\delta A_{\mu} = c_{\mu}  {\bf 1},
\label{Tsym}
\end{equation}
whose transformation is
proportional to the unit matrix ${\bf 1}$
and an extra supersymmetry,
\beq
\cases{
\delta^{(2)} A_\mu & = 0  \cr
\delta^{(2)} \psi & $=\epsilon_2$.
}
\label{Ssym2}
\eeq
The linear combinations of  these two supersymmetries (\ref{Ssym1}) and
(\ref{Ssym2}),
\begin{equation}
\tilde{Q}^{(1)} = Q^{(1)} + Q^{(2)}, \
\tilde{Q}^{(2)} = i( Q^{(1)} - Q^{(2)}),
\end{equation}
 satisfy the following commutation relations,
\begin{equation}
[\bar{\epsilon_1} \tilde{Q^{(i)}}, \bar{\epsilon_2} \tilde{Q^{(j)}}]
= -2 \bar{\epsilon_1} \gamma_{\mu} \epsilon_2 p^{\mu} \delta^{(ij)},
\end{equation}
where $p^{\mu}$ is the generator of the translation (\ref{Tsym})
and $i,j= 1, 2$.
Therefore, if we  interpret  the eigenvalues of the  matrices $A_{\mu}$
as our space-time coordinates,
the above symmetries
can be regarded as  ten-dimensional ${\cal N}=2$ space-time
supersymmetry.
Since the maximal space-time supersymmetry  guarantees  the existence of
gravitons if the theory permits the massles spectrum, it supports our 
conjecture  that the IIB matrix model
is a constructive definition of superstring.
This is one of the major reasons to interpret the eigenvalues of
$A_{\mu}$ as being  the coordinates of the space-time which has emerged
out of the matrices.
\par
The second important and confusing  property
is that the model has the same
action as  the low-energy effective action of D-instantons.
\cite{Witten}
We should emphasize here  the differences between these two theories,
since we are led to different interpretations of space-time.
From an effective-theory point of view,  the eigenvalues represent
the coordinates of D-instantons in
ten-dimensional  bulk space-time,  which we have assumed a priori
from the beginning of constructing the effective action.
On the other hand,
from a  constructive point of view, we cannot assume such a
bulk space-time, in which matrices live.
This is because
not only fields, but also the space-time,
should be  dynamically generated as a result of the
dynamics of the matrices.
The space-time  should be constructed only from the  matrices.
The most natural interpretation is that  space-time consists of
$N$ discretized  points, and  that the eigenvalues represent their
space-time coordinates.
Here we need to assume that the dynamics of the IIB matrix model is such that
the resulting eigenvalue distributions are smooth enough to be interpreted as
Riemannian geometry.
\par
A final important property is that the type-IIB matrix model has no
free parameters. The coupling constant $g$ can  always be absorbed by
field redefinitions:
\begin{eqnarray}
\cases{
A_{\mu} \rightarrow g^{1/2} A_{\mu} \cr
\psi \rightarrow g^{3/4} \psi.
}
\end{eqnarray}
This is reminiscent of  string theory where a  shift of the string coupling
constant is always absorbed to that of the  dilaton vacuum expectation
value (vev).
In an  analysis of the Schwinger-Dyson equation
of the IIB matrix model, \cite{FKKT}
we introduced an infrared cut-off $\epsilon$,  which gives a string
coupling constant, $g_{\rm st} = 1/N\epsilon^2$.
However, through a more careful analysis
of  the dynamics of the eigenvalues, \cite{AIKKT}  we have shown that there is
no such  infrared divergences associated with  infinitely
separated eigenvalues, and that
the infrared cutoff $\epsilon$ which we have
introduced by hand can be
determined dynamically in terms of $N$ and $g$.
(The Schwinger-Dyson equation and the double scaling limit are
discussed in \S2.)
\par
We then explain  several reasons
we believe that the IIB matrix model is a
constructive definition of type-IIB superstring
in addition to the symmetry argument.
First,
this action can be related to the Green-Schwarz action of a
superstring\cite{GS}
by using  the semiclassical correspondence in the large $N$ limit:
\beqa
-i[\;,\;] &\rightarrow&  \{\;,\;\}, \n
{\rm Tr} &\rightarrow&  \int {d^2 \sigma }\sqrt{\hat{g}} .
\label{correspondence}
\eeqa
In fact, Eq. (\ref{action}) is related to the Green-Schwarz action
in the Schild gauge:\cite{Schild}
\beq
S_{\rm Schild}=\int d^2\sigma \left[\sqrt{\hat{g}}\alpha\left(
\frac{1}{4}\{X^{\mu},X^{\nu}\}^2
-\frac{i}{2}\bar{\psi}\Gamma^{\mu}\{X^{\mu},\psi\}\right)
+\beta \sqrt{\hat{g}}\right].
\label{SSchild}
\eeq
We need to integrate over the scale factor of the metric $\sqrt{g}$
in order to quantize the Schild action
\beq
Z=\int D\sqrt{g}DXD\psi e^{-S_{\rm Schild}}.
\eeq
The matrix analog is the following grand canonical ensemble:
\beqa
Z&=&\sum_{n=0}^{\infty}\int dAd\psi e^{-S(\beta )},\n\
S(\beta )&=&\alpha \left(-{1\over 4}{\rm Tr}[A_{\mu},A_{\nu}]^2-{1\over
2}{\rm Tr}(\bar{\psi}\Gamma^{\mu}[A_{\mu},\psi])\right) +\beta {\rm Tr} 1 .
\eeqa
If the large $N$ limit is smooth, we expect that the $\beta \rightarrow
\beta_C$
limit is identical to consider the microcanonical ensemble with
fixed $N$ and take $N$ large.
\par
The correspondence can go farther
beyond the above identification of the  model with  a matrix regularization
of the first quantized superstring.
Namely, we can describe an arbitrary  number of interacting D-strings
and anti-D-strings as blocks of matrices, each of which corresponds
to the matrix regularization of a string.
Off-diagonal blocks induce interactions between  these strings.
\cite{IKKT,effective}
Thus, it must be clear that the IIB matrix model is definitely not the first
quantized theory of a D-string, but a full second quantized theory.
\par
It has also been  shown\cite{FKKT} that  Wilson loops satisfy  the
string field equations of motion for type-IIB superstring in the
light-cone gauge, which is a second evidence for the conjecture
that the IIB matrix model is a constructive definition of superstring.
We consider the following regularized Wilson loop: \cite{FKKT,hamada}
\beq
w(C)  =  {\rm Tr}\left[\prod_{n=1}^M {\rm exp}\{ i\epsilon (k_{n}^{\mu} A_{\mu}
+\bar{\lambda}_{n} \psi ) \}\right ] .
\label{Wilsonloop-intro}
\eeq
Here, $k_n^{\mu}$ are the momentum densities distributed along a loop $C$;
we have also introduced  fermionic sources,
$\lambda _{n}$.
The symbol $\epsilon$
is a short-distance cutoff of string world sheet.
In the large $N$ limit,
$\epsilon$ should  go to $0$ so as to satisfy the double scaling limit.
In Ref. \citen{FKKT} it was proved, once we have taken the
correct scaling limit,  that the ${\cal N}=2$
supersymmetry is enough to reproduce the lightcone field equation of
type IIB superstring from the IIB matrix model.
In order to resolve the problem of the double scaling limit, we need to
evaluate several quantities (i.e., an expectation value of the
Wilson loop with almost zero total momentum) and
there is still some subtlety as to how to take
this double scaling limit in which  we obtain an interacting
string theory. It is discussed extensively in \S2.
\par
Considered as a matrix regularization of the Green-Schwarz IIB superstring,
the IIB matrix model describes interacting D-strings. On the other hand,
in an analysis of  the  Wilson loops, the IIB matrix model describes
joining and splitting interactions of fundamental IIB superstrings
created by the Wilson loops.
From these considerations, it is plausible to conclude
that if we can take the
correct double scaling limit, the  IIB matrix model could
become  a constructive
definition of type-IIB superstring. Furthermore, we believe that
all string theories are connected by duality transformations, and once
we  construct a nonperturbative  definition of any one of them,
 we can
describe the vacua of any other strings, particularly the true vacuum in
which we live.
\par
The dynamics of eigenvalues, that is, dynamical generation
of  space-time  was first discussed in Ref. \citen{AIKKT}.
An effective action of  eigenvalues can be
obtained by integrating all of
the off-diagonal bosonic and fermionic components,
and then the diagonal fermionic coordinates (which we call fermion zeromodes).
If we quench the bosonic diagonal components,
$x_{\mu}^{i}$ ($i=1\cdots N$),
and neglect the
fermion zeromodes $\xi^{i}$, the effective action for $x_{\mu}^{i}$
coincides with that of
supersymmetric Yang-Mills theory with maximal supersymmetry and vanishes
respecting  the stability of the  supersymmetric moduli.
The inclusion of fermion zeromodes as well as the
 non-planar contributions
lifts the degeneracy, and we can obtain a nontrivial effective action
for the space-time dynamics.
In Ref. \citen{AIKKT} we estimated this effective action
by perturbation at one loop, which is valid when all eigenvalues
are far from one another,  $|{\boldsymbol x}_i-{\boldsymbol x}_j| 
\gg \sqrt{g}$.
Of course, this one-loop effective action is not sufficient to
determine the full space-time structure,  but we expect
that it captures
some of the essential points concerning
the  formation of  space-time.
One of  important properties of the effective action is that,
as a result of  grassmannian integration of the fermion zeromodes,
 space-time points make a network connected locally by bond interactions.
This feature becomes important when we extract diffeomorphism symmetry
from our matrix model, which is discussed in \S4.
\par
Once we are convinced that IIB matrix model is a constructive
definition of superstring, we then have to give natural
interpretation of low energy dynamics. That is, we need to show
how we can obtain local field theory in a low energy approximation
and the origin of local gauge symmetry in our space-time generated
dynamically from matrices.
We also have to show how the background metric is encoded
in a low energy field theory in the space-time, especially
the origin of diffeomorphism invariance.
In Ref.~\citen{IK} we have shown that,
 if we suppose  that the  eigenvalue
distribution consists of small clusters of size $n$,
the low-energy theory acquires $SU(n)$ local space-time gauge symmetry.
This gauge invariance assures the existence of
a gauge field  propagating
in the  space-time of  distributed eigenvalues.
Also, we have obtained a low energy effective action
(a gauge-invariant kinetic action) for a
fermion in the adjoint representation of $SU(n)$, which becomes massless.
The low-energy behavior  for these fields is formulated
as   a lattice gauge theory on a dynamically generated random lattice,
and hence supports our interpretation of  space-time.
We have also shown in Ref. \citen{IK} that
the diffeomorphism invariance of our model  originates in
 invariance under  permutations  of the eigenvalues.
Our model realizes the invariance in an interesting way by
summing all possible graphs connecting the space-time points.
The diffeomorphism invariance  restricts  the  low-energy behavior
of the model,
and indicates the existence of a massless graviton in the low energy
effective field theory on dynamically generated space-time.
The background metric for propagating  fields is shown to be
encoded in the density
correlation of the  eigenvalues, while the dilaton vacuum expectation value
is encoded in the eigenvalue density.
A curved background can be described as a nontrivial distribution
of eigenvalues whose density correlation behaves inhomogeneously.
\par
Both of these fundamental symmetries, local gauge symmetry and the
diffeomorphism symmetry, originates in the $SU(N)$ invariance of the
matrix model.
It is  quite interesting  that these fundamental symmetries can arise from
a very simple matrix model defined on a single point.
These are discussed in \S4.
\par
The organization of this paper is as follows. In \S2,
we review our analysis of the Schwinger-Dyson equation for the
Wilson loops and discuss a problem on the double scaling limit.
In \S3, we
briefly review our analysis on  the dynamics of  space-time,
and show how a  network picture of  space-time arises.
Here is an analogy with the dynamical triangulation approach to
quantum gravity.
We also discuss a  recent result  on  numerical simulation.
In \S4, we discuss a possible origin of low energy
gauge symmetry on a dynamically generated space-time and that of
the diffeomorphism invariance.
We also show that we can obtain a low energy effective action
for several fields by using these low energy symmetries.
Such  low-energy effective theories  are  formulated as
a lattice gauge theory on a dynamically generated random lattice.
Section 5 is devoted to  discussions.
\section{Loop equations and scaling limit}
\setcounter{equation}{0}
In this section, we derive
the light-cone string field theory of type IIB
super-\linebreak
string\cite{GSB}
from the Schwinger-Dyson equations
for the Wilson loops (loop equations).
The purpose of this analysis is
to verify that the IIB matrix
model indeed reproduces the standard
perturbation series of type IIB
superstring and to
fix how to take the scaling limit.
Here we review the analysis in Ref. \citen{FKKT} with refinement.

We regard the Wilson loop
\beqa
w(C)&=&{\rm Tr}(v(C)),\n
v(C)&=&P \exp \left(i\int_{0}^{2\pi} d\sigma (k^{\mu}(\sigma) A_{\mu}
+\bar{\lambda}(\sigma)\psi)\right ),
\label{Wilsonloop}
\eeqa
as the creation or annihilation operator for the momentum
representation eigenstate of string $|k^{\mu},\lambda(\sigma)\rangle$,
where $k^{\mu}(\sigma)$ is a momentum density on the worldsheet and
$\lambda(\sigma)$ is its super-partner. We explain in \S2.1 the reason
why this interpretation is natural.

The basic equations we consider in the following
are the loop equations:
\beq
0=\int dAd\psi
{\partial \over \partial A_{\mu}^\alpha}
\{{\rm Tr}
(t^\alpha v(C_1))w(C_2)\cdots w(C_l)e^{-S}\} ,
\label{eqA}
\eeq
\beq
0=\int dAd\psi
{\partial \over \partial \psi ^\alpha }
\{{\rm Tr}(t^\alpha v(C_1))w(C_2)\cdots w(C_l)e^{-S}\} ,
\label{eqpsi}
\eeq
where $t^{\alpha}$ is a generator of $U(N)$ Lie algebra,
and an equation which represents
the local reparametrization invariance of the loop:
\beq
\left(k^{\mu}(\sigma) \left(\frac{\delta}{\delta k^{\mu}(\sigma)}\right)' +
\bar{\lambda}(\sigma)\left(\frac{\delta}{\delta \bar{\lambda}(\sigma)}\right)'\right)w(C)=0.
\label{localrepinv}
\eeq

\subsection{Wilson loops and light-cone setting}

In this subsection, we briefly sketch our basic idea
for deriving the light-cone string field theory from
the loop equations.
For this purpose, let us consider
only the bosonic parts.
We emphasize here that we perform this simplification
for explanation.
In fact, as is explained later, we cannot
obtain the light-cone string field theory from the bosonic
reduced model.

We first explain a motivation to consider the Wilson loops
like Eq.~(\ref{Wilsonloop}) in the following.
Let us consider a gauge theory in a box with size $a$.
We impose the
periodic boundary conditions on the fields.
Then a Wilson (Polyakov) loop is given by
\beq
w(C)={\rm Tr}\left(Pe^{i\int_{0}^{2\pi}
d\sigma x^{'\mu}(\sigma) A_{\mu}(x^{\mu}(\sigma))}\right),
\label{finitevolumeWilsonloop}
\eeq
where $x^{\mu}$ is an arbitrary function which satisfies
the condition
\beq
x^{\mu}(2\pi)-x^{\mu}(0)=n^{\mu}a.
\eeq
Here the $n^{\mu}$ are the winding numbers in the $\mu$-th directions.
In the zero volume limit ($a\rightarrow 0$),
Eq. (\ref{finitevolumeWilsonloop}) reduces to
\beq
w(C)={\rm Tr}\left(Pe^{i\int_{0}^{2\pi} d\sigma x^{'\mu}(\sigma) A_{\mu}}\right).
\label{Wilsona}
\eeq
The expectation values of the Wilson loops with nontrivial windings
vanish if the translation ($U(1)$) symmetry is not spontaneously broken.
This phase transition is known to be the deconfining transition
in lattice gauge theory.
This symmetry may be broken in the large $N$ limit for bosonic reduced models.
In order to obtain string theory in Minkowski space which is
translation invariant, we have to keep the translation invariance
by supersymmetrizing the theory.
We assume that the gauge theory is in the confining phase and
hence well described by string theory.
We further assume that
there is no phase transition while we take $a \rightarrow 0$ limit.
Then the Wilson loop (\ref{Wilsona}) must represent strings.
Let us consider the Wilson loops with large winding numbers so that
$n^{\mu}a$ is finite in $a\rightarrow0$ limit.
They represent the strings with no momentum
but with many windings
in the zero volume target space.
In order to obtain the strings
moving in the infinite-volume target space,
we adopt the T-dual picture
here. That is, we reinterpret $x'^{\mu}$
as the momentum density, $k^{\mu}$,
and obtain the (bosonic part of) expression (\ref{Wilsonloop}).
As is expected in the ordinary T-dual picture, the windings
are converted to the total momenta, which is seen
readily in the relation
\beq
n^{\mu}a=\int^{2\pi}_{0}d\sigma k^{\mu}(\sigma).
\eeq
Thus we regard the Wilson loop (\ref{Wilsonloop})
as the creation and
annihilation operator
for the momentum representation eigenstate of string
$|k^{\mu}(\sigma), \lambda(\sigma)\rangle$.
Since $A_{\mu}$ is dual to
the momentum in the
expression (\ref{Wilsonloop}), it can be interpreted as
the space-time coordinate naturally.

\begin{figure}[b]
\epsfxsize=8cm
\centerline{\epsfbox{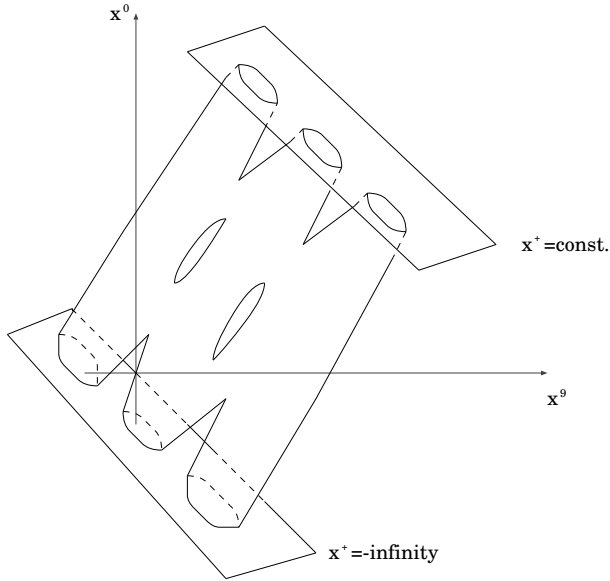}}
\caption{The light-cone setting. We consider the configurations of
the Wilson loops which possess the identical light-cone time $x^{+}$.
We also locate a group of the Wilson loops at $x^{+}=-\infty$ which represent
a particular initial state. We put $k^{+}(\sigma)=1$
for all the Wison loops by using the reparametrization invariance.}
\label{lightcone}
\end{figure}

To make the connection with the light-cone string field theory, we
consider the particular configurations of the Wilson loops,
which we call the light-cone setting. The Schwinger-Dyson equations lead to
the continuum loop equation as we explain shortly:
\beq
(k^{(b)\mu}(\sigma)^2 + x'^{(b)\mu}(\sigma)^2)
\langle w(C_1)\cdots w(C_b) \cdots w(C_l)\rangle=0,
\label{Virasoro}
\eeq
where for simplicity we consider only the free part and
we denote $i \frac{\delta}{\delta k^{\mu} (\sigma)}$ as
$x_{\mu}(\sigma)$.
We have also the local reparametrization invariance
(the bosonic part of (\ref{localrepinv})),
\beq
x'^{(b)\mu}(\sigma)k^{(b)}_{\mu}(\sigma)
\langle w(C_1)\cdots w(C_b) \cdots w(C_l)\rangle=0 .
\label{repinv}
\eeq
We put $k^{(b)+}(\sigma)=1$ for all the Wison loops
by using the reparametrization invariance so that
we set the length of the strings to be equal to
the $+$ components of their total momenta.
We consider the configurations of the Wilson loops
which possess the identical
light-cone time $x^+$.
Namely we perform the functional Fourier transformations of the
Wilson loop from $k^{(b)-}(\sigma)$ to $x^{(b)+}(\sigma)$ and
consider such configurations that
$x^{(b)+}(\sigma)=x^{+}$ for all the Wilson loops.
We also locate a group of the Wilson loops
at $x^+ =-\infty$ which represent a particular initial state.
After these prescriptions, we denote the Wilson loop by
$\tilde{w}(C)$:
\beqa
\tilde{w}(C)&=&\tilde{w}[x^+(\sigma)=x^+,k^+(\sigma)=1,k^i(\sigma)
;\sigma= 0 \sim p^+] \n
&=&\int Dk^- (\sigma) e^{-i\int_{0}^{p^+}d\sigma
x^+ (\sigma)k^- (\sigma)}w(C) |_{
k^+(\sigma)=1,
x^+(\sigma)=x^+},
\eeqa
where $p^+$ is the $+$ component of the total momentum.
This is the light-cone setting which is illustrated
in Fig. \ref{lightcone}.

In the light-cone setting, the loop equation
(\ref{Virasoro}) reduces to
\beqa
&&\hspace{-1cm}k^{(b)-}(\sigma)
\langle \tilde{w}(C_1)\cdots \tilde{w}(C_b) \cdots
\tilde{w}(C_l)\rangle\n
&&=i\frac{\delta}{\delta x^{(b)+}(\sigma)}
\langle \tilde{w}(C_1)\cdots \tilde{w}(C_b) \cdots
\tilde{w}(C_l) \rangle\n
&&=\frac{1}{2}(k^{(b)i}(\sigma)^2 + x'^{(b)i}(\sigma)^2)
\langle \tilde{w}(C_1)\cdots \tilde{w}(C_b) \cdots
\tilde{w}(C_l)\rangle.
\label{k-}
\eeqa
Equation (\ref{repinv}) reduces to
\beqa
&&\hspace{-1cm}x'^{(b)-}(\sigma)
\langle \tilde{w}(C_1)\cdots \tilde{w}(C_b) \cdots
\tilde{w}(C_l)\rangle\n
&&=-i\left(\frac{\delta}{\delta k^{(b)+}(\sigma)}\right)'
\langle \tilde{w}(C_1)\cdots \tilde{w}(C_b) \cdots
\tilde{w}(C_l)\rangle\n
&&=x'^{(b)i}(\sigma)k_{(b)i}(\sigma)
\langle \tilde{w}(C_1)\cdots \tilde{w}(C_b) \cdots
\tilde{w}(C_l)\rangle.
\eeqa
Using these equations we can deform the Wilson loop
locally from the constant $x^+$ surface and shift
the value of $k^{(b)+}(\sigma)$ locally. Therefore we can
recover general configurations of the Wilson loops.
This fact ensures us to consider the light-cone setting.
Integrating Eq.
(\ref{k-}) over $\sigma$ and summing up over $b$,
we obtain an operator which shift the constant value of $x^+$,
that is, the light-cone Hamiltonian:
\beqa
&&\hspace{-1cm}i\frac{\partial}{\partial x^{+}}
\langle \tilde{w}(C_1)\cdots \tilde{w}(C_b) \cdots
\tilde{w}(C_l)\rangle\n
&&=\sum_{b} \int d \sigma \frac{1}{2} (k^{(b)i}(\sigma)^2
+ x'^{(b)i}(\sigma)^2)
\langle \tilde{w}(C_1)\cdots \tilde{w}(C_b) \cdots
\tilde{w}(C_l)\rangle,
\eeqa
which is identical to the free light-cone
Hamiltonian of bosonic string.
Note that we can obtain a closed system of equations within
the light-cone setting even though we consider the particular
configurations of the Wilson loops.
Our procedure is analogous to obtaining a Hamiltonian
in super-many-time theory,
where a Hamiltonian density is
defined on a general space-like surface.
We consider a constant time surface and integrate a Hamiltonian density on it
to obtain an ordinary Hamiltonian.

\subsection{Loop equations and loop space}
In this subsection, we represent our loop equations
by the loop space variables and put them in the light-cone
setting.

The loop equation (\ref{eqA}) is evaluated as
\beqa
0&=&\frac{1}{g^2} \left\langle
({\rm Tr}\left([A_{\nu},[A^{\mu},A^{\nu}]]
+\frac{1}{2}\{\bar{\psi}\Gamma^{\mu},\psi \}\right)
v(C_1)) w(C_2) \cdots w(C_l)
\right\rangle \n
&&+ i\int_{0}^{2\pi} d\sigma  k_{\mu}^{(1)}(\sigma)
\big\langle
{\rm Tr}\big(P e^{i\int_{0}^{\sigma}d\sigma'
(k^{(1)\mu}(\sigma')A_{\mu}+\bar{\lambda}^{(1)}(\sigma')\psi)}\big)\nonumber\\
&&\times {\rm Tr}\big(P e^{i\int_{\sigma}^{2\pi}d\sigma''
(k^{(1)\mu}(\sigma'')A_{\mu}+\bar{\lambda}^{(1)}(\sigma'')\psi)}\big)
w(C_2) \cdots w(C_l)\big \rangle \n   
&&+ i\sum_{b=2}^{l}
\int_{0}^{2\pi} d\sigma  k_{\mu}^{(b)}(\sigma)
\big\langle
{\rm Tr}\big(P e^{i\int_{\sigma}^{2\pi}d\sigma'
(k^{(b)\mu}(\sigma')A_{\mu}+\bar{\lambda}^{(b)}(\sigma')\psi)}
v(C_1)\nonumber\\
&&\times P e^{i\int_{0}^{\sigma}d\sigma''
(k^{(b)\mu}(\sigma'')A_{\mu}+\bar{\lambda}^{(b)}(\sigma'')\psi)}\big)
w(C_2) \cdots \check{w}(C_b)  \cdots w(C_l)\big \rangle,
\label{eqA2}
\eeqa
where $\check{w}(C_b)$ implies the absence of the Wilson loop
$w(C_b)$.
The first term in Eq.~(\ref{eqA2}) represents the infinitesimal deformation
of the string coming from the variation of the action $S$ while
the second and third terms represent the splitting and joining interaction
of strings respectively.
Similarly, the loop equation (\ref{eqpsi}) is evaluated as
\beqa
0&=&\frac{1}{g^2} \langle
({\rm Tr}(\Gamma^{\mu}[A_{\mu},\psi]
v(C_1)) w(C_2) \cdots w(C_l)
\rangle \nonumber\\
&&+ i\int_{0}^{2\pi} d\sigma  \lambda^{(1)}(\sigma)
\big\langle
{\rm Tr}\big(P e^{i\int_{0}^{\sigma}d\sigma'
(k^{(1)\mu}(\sigma')A_{\mu}+\bar{\lambda}^{(1)}(\sigma')\psi)}\big)\nonumber\\
&&\times {\rm Tr}\big(P e^{i\int_{\sigma}^{2\pi}d\sigma''
(k^{(1)\mu}(\sigma'')A_{\mu}+\bar{\lambda}^{(1)}(\sigma'')\psi)}\big)
 w(C_2) \cdots w(C_l)\big \rangle \n
&&+ i\sum_{b=2}^{l}
\int_{0}^{2\pi} d\sigma  \lambda^{(b)}(\sigma)
\big\langle
{\rm Tr}\big(P e^{i\int_{\sigma}^{2\pi}d\sigma'
(k^{(b)\mu}(\sigma')A_{\mu}+\bar{\lambda}^{(b)}(\sigma')\psi)}
v(C_1)\nonumber\\
&&\times P e^{i\int_{0}^{\sigma}d\sigma''
(k^{(b)\mu}(\sigma'')A_{\mu}+\bar{\lambda}^{(b)}(\sigma'')\psi)}\big)
w(C_2) \cdots \check{w}(C_b)  \cdots w(C_l) \big\rangle.
\label{eqpsi2}
\eeqa

We need to represent these equations as differential operators on the
loop space. Then we treat field insertions in the loop such as
\beq
{\rm Tr}([A_{\mu},\psi]v(C)).
\eeq
We have the following identity,
\beqa
&&\hspace{-1cm}{\rm Tr}([A_{\mu},\psi]v(C))\n
&=&-\left(\frac{\delta}{\delta k^{\mu}(\frac{\ep}{2})}
-\frac{\delta}{\delta k^{\mu}(-\frac{\ep}{2})}\right)
\frac{\delta}{\delta \bar{\lambda}(0)}w(C)\n
&&-\ep\left(\frac{\delta}{\delta k^{\mu}(\frac{\ep}{2})}
-\frac{\delta}{\delta k^{\mu}(-\frac{\ep}{2})}\right)
\frac{\delta}{\delta \bar{\lambda}(0)}
\left(k^{\nu}(0)\frac{\delta}{\delta k^{\nu}(0)}
+\bar{\lambda}(0)\frac{\delta}{\delta \bar{\lambda}(0)}\right)w(C)\n
&&+\mbox{higher order of $\ep$}.
\label{commutator}
\eeqa
Though the right-hand side of Eq.~(\ref{commutator}) vanishes
in naive $\ep \rightarrow 0$ limit,
we expect here that for finite $N$ an ultraviolet cut-off $\ep$
appears naturally on the worldsheet. We assume that $\ep$
converges to zero in $N \rightarrow \infty$ limit such as
$\ep \sim N^{-b} \; (b > 0)$.
We should keep $\ep$ and other quantities depending on
$N$ in the process of the calculation and
take $N \rightarrow \infty \; (\ep \rightarrow 0)$ limit
as  the continuum limit in the final stage.

We expect naively that we may ignore the terms except the first one
in the right-hand sides of Eq.~(\ref{commutator}) since
they are the higher order terms with respect to $\ep$.
However, in general,
they contribute to the renormalization
of the lower-dimensional terms
because in the loop equations they generate
the divergences from operator product expansion. The ways
we represent the loop equations
are not unique since the right-hand sides
of Eq.~(\ref{commutator}) can be expressed in infinitely many
ways. However we expect that they are unique
in the $\ep\rightarrow 0$ limit in the loop equation.
This is the universality of the differential operators, which is guaranteed by
a power counting and symmetry as seen in \S 2.4.
Here we may draw an analogy with the
quantum field theory on the lattice.
The lattice action may be expanded formally in terms of the lattice
spacing $a$.
Although the operators which are suppressed by the powers of $a$
formally vanish in the continuum, we cannot simply neglect them
because they may renormalize the relevant operators.
In fact we can write down many lattice actions which possess the identical
continuum limit.

In the following, we often show only the naive leading
terms in the loop equations.
Note that we cannot, in fact,  neglect infinitely many
higher order terms of $\ep$ as is discussed above.

By multiplying  Eq. (\ref{eqA2}) by $-i k_{\mu}^{(1)}(0)$,
we obtain
\beqa
0&=&\frac{1}{g^2}
\bigg\langle\bigg(-\frac{1}{\epsilon}
\left(\frac{\delta}{\delta k^{(1)\mu}
(\frac{\epsilon}{2})}-
\frac{\delta}{\delta k^{(1)\mu}
(-\frac{\epsilon}{2})}\right)^2\nonumber\\
&&-i k^{(1)}_{\mu}(0)\left(\frac{\delta}{\delta \bar{\lambda}^{(1)}
(\frac{\epsilon}{2})}
-\frac{\delta}{\delta \bar{\lambda}^{(1)}(-\frac{\epsilon}{2})}\right)
\Gamma^0 \Gamma^{\mu}
\frac{\delta}{\delta \bar{\lambda}^{(1)}(-\frac{\epsilon}{2})}\bigg)
w(C_1) \cdots w(C_l)\bigg
\rangle \nonumber\\
&&+ \int_{0}^{2\pi} d\sigma k^{(1)\mu}(0) k_{\mu}^{(1)}(\sigma)
\big\langle
{\rm Tr}\big(P e^{i\int_{0}^{\sigma}d\sigma'
(k^{(1)\mu}(\sigma')A_{\mu}+\bar{\lambda}^{(1)}(\sigma')\psi)}\big)\nonumber\\
&&\times {\rm Tr}\big(P e^{i\int_{\sigma}^{2\pi}d\sigma''
(k^{(1)\mu}(\sigma'')A_{\mu}+\bar{\lambda}^{(1)}(\sigma'')\psi)}\big)
w(C_2) \cdots w(C_l)\big \rangle \n
&&+ \sum_{b=2}^{l}
\int_{0}^{2\pi} d\sigma k^{(1)\mu}(0) k_{\mu}^{(b)}(\sigma)
\big\langle
{\rm Tr}\big(P e^{i\int_{0}^{\sigma}d\sigma'
(k^{(b)\mu}(\sigma')A_{\mu}+\bar{\lambda}^{(b)}(\sigma')\psi)}\n
&&\times v(C_1)
P e^{i\int_{\sigma}^{2\pi}d\sigma''
(k^{(b)\mu}(\sigma'')A_{\mu}+\bar{\lambda}^{(b)}(\sigma'')\psi)}\big)
w(C_2) \cdots \check{w}(C_b)  \cdots w(C_l)\big \rangle.\nonumber\\
\label{eqA3}
\eeqa
Similarly, Eq. (\ref{eqpsi2}) is represented as
\beqa
0&=&\frac{i}{g^2}
\left\langle
\left(\frac{\delta}{\delta k^{(1)\mu}
(\frac{\epsilon}{2})}-
\frac{\delta}{\delta k^{(1)\mu}
(-\frac{\epsilon}{2})}\right)\Gamma^{\mu}
\frac{\delta}{\delta \bar{\lambda}^{(1)}(0)}
w(C_1) \cdots w(C_l)
\right \rangle \n
&&+ \int_{0}^{2\pi} d\sigma \lambda^{(1)}(\sigma)
\big\langle
{\rm Tr}\big(P e^{i\int_{0}^{\sigma}d\sigma'
(k^{(1)\mu}(\sigma')A_{\mu}+\bar{\lambda}^{(1)}(\sigma')\psi)}\big)\nonumber\\
&&\times {\rm Tr}\big(P e^{i\int_{\sigma}^{2\pi}d\sigma''
(k^{(1)\mu}(\sigma'')A_{\mu}+\bar{\lambda}^{(1)}(\sigma'')\psi)}\big)
w(C_2) \cdots w(C_l) \big\rangle \n
&&+ \sum_{b=2}^{l}
\int_{0}^{2\pi} d\sigma \lambda^{(b)}(\sigma)
\big\langle
{\rm Tr}\big(P e^{i\int_{\sigma}^{2\pi}d\sigma'
(k^{(b)\mu}(\sigma')A_{\mu}+\bar{\lambda}^{(b)}(\sigma')\psi)}\nonumber\\
&&\times v(C_1)
P e^{i\int_{0}^{\sigma}d\sigma''
(k^{(b)\mu}(\sigma'')A_{\mu}+\bar{\lambda}^{(b)}(\sigma'')\psi)}\big)
w(C_2) \cdots \check{w}(C_b)  \cdots w(C_l) \big\rangle.\n
\label{eqpsi3}
\eeqa

Here let us proceed to the light-cone setting.
The $+$ component of total momentum carried by the
$\tilde{w}[x^{+}(\sigma')=x^{+},k^{+}(\sigma')=1
,k^i(\sigma'),\lambda(\sigma');\sigma'=0 \sim \sigma)]$
is equal to $\sigma$.
Therefore we expect naively
\beq
\langle\tilde{w}[x^{+}(\sigma')=x^{+},k^{+}(\sigma')=1
,k^i(\sigma'),\lambda(\sigma);\sigma'=0 \sim \sigma)]
\rangle=0
\eeq
because of
the momentum conservation if $\sigma \neq 0$.
We also expect the same case for
$\langle\tilde{w}[x^{+}(\sigma')=x^{+},k^{+}(\sigma')=1
,k^i(\sigma'),\lambda(\sigma');
\sigma'=p^+ -\sigma \sim p^+]\rangle$.
However these are not actually the cases
since the eigenvalues of $A_{\mu}$ distribute
in a finite range for finite $N$, which violates the momentum conservation
slightly.
Therefore these two $\tilde{w}$ have support for small
$\sigma$ generally.
We define a nonvanishing quantity $I$ by
\beqa
I&=&\int_{0}^{p^+}d\sigma
 \langle\tilde{w}[x^{+}(\sigma')=x^{+},k^{+}(\sigma')=1,
k^i(\sigma'),\lambda(\sigma);\sigma'=0 \sim \sigma]\n
&&+\tilde{w}[x^{+}(\sigma')=x^{+},k^{+}(\sigma')=1,
k^i(\sigma'),\lambda(\sigma);\sigma'=p^+-\sigma \sim p^+]\rangle.
\eeqa
We expect that $I$ diverges as $N \rightarrow \infty$ and
assume that
its large $N$ behavior is $I \sim N^a \; (a > 0)$.
Then a part of the splitting interaction term in the loop equations
contribute to the first term in the loop equations representing
the infinitesimal deformation of the string.

Thus in the light-cone setting Eq. (\ref{eqA3}) leads to
\beqa
&&\hspace*{-1cm} k^{(1)-}(0) \langle \tilde{w}(C_1) \cdots \tilde{w}(C_l) \rangle\n
&=&\frac{1}{Z}\bigg \{
\bigg(k^{(1)i}(0)^2+\frac{\ep}{g^2 I}x^{'(1)}(0)^2 \n
&&-i \sqrt{2} \frac{\ep}{g^2 I} \theta^{'(1)}_{a}(0)\theta^{(1)}_{a}(0)
-i \frac{\ep}{g^2 I}k^{(1)i}(0)\gamma^{i}_{a \da}
(\theta^{'(1)}_{a}(0)\theta^{(1)}_{\da}(0)
+\theta^{'(1)}_{\da}(0)\theta^{(1)}_{a}(0))\bigg)\n
&&\hspace*{5cm} \times \langle \tilde{w}(C_1) \cdots \tilde{w}(C_l) \rangle\n
&&+\frac{1}{I} \int_{0}^{p^{(1)+}} d\sigma
k^{(1)}_{\mu}(0) k^{(1)\mu}(\sigma)
\big\langle
\tilde{\rm Tr}\big(P e^{i\int_{0}^{\sigma}d\sigma'
(k^{(1)\mu}(\sigma')A_{\mu}+\bar{\lambda}^{(1)}(\sigma')\psi)}\big)\nonumber\\
&&\times \tilde{\rm Tr}\big(P e^{i\int_{\sigma}^{p^{(1)+}}d\sigma''
(k^{(1)\mu}(\sigma'')A_{\mu}+\bar{\lambda}^{(1)}(\sigma'')\psi)}\big)
\tilde{w}(C_2) \cdots \tilde{w}(C_l)\big \rangle \n
&&+ \frac{1}{I}\sum_{b=2}^{l}
\int_{0}^{p^{(b)+}} d\sigma
k^{(1)}_{\mu}(0) k^{(b)\mu}(\sigma)
\big\langle
\tilde{\rm Tr}\big(P e^{i\int_{\sigma}^{p^{(b)+}}d\sigma'
(k^{(b)\mu}(\sigma')A_{\mu}+\bar{\lambda}^{(b)}(\sigma')\psi)} \n
&&\times v(C_1)
P e^{i\int_{0}^{\sigma}d\sigma''
(k^{(b)\mu}(\sigma'')A_{\mu}+\bar{\lambda}^{(b)}(\sigma'')\psi)}\big)
\tilde{w}(C_2) \cdots \check{\tilde{w}}(C_b)  \cdots \tilde{w}(C_l)\big \rangle \bigg\}.\nonumber\\
\label{eqk-}
\eeqa
where $Z$ is defined by
\beq
Z=2\left(1+\frac{i \ep}{\sqrt{2} g^2 I} \theta'_{\dot{a}}(0)
\theta_{\dot{a}}(0)\right)
\eeq
and $\tilde{\rm Tr}$ means that the corresponding Wilson loop
is arranged in the light-cone setting.
Here $x_{\mu}$, $\theta_{a}$ and $\theta_{\da}$ are
defined as follows:
\beqa
x_{\mu}(\sigma)=i \frac{\delta}{\delta k^{\mu}(\sigma)},\n
\theta_{a}(\sigma)=\frac{\delta}{\delta \lambda_{a}(\sigma)},\n
\theta_{\da}(\sigma)=\frac{\delta}{\delta \lambda_{\da}(\sigma)}.
\eeqa

Equation (\ref{eqpsi3}) is decomposed into two equations as follows.
\beqa
&&\hspace*{-1cm}\lambda^{(1)}_{\dot{a}}(0)
\langle \tilde{w}(C_1) \cdots \tilde{w}(C_l) \rangle\n
&=&\frac{\ep}{g^2 I}
(\sqrt{2} x^{'(1)-}(0) \theta_{\da}(0) + x^{'i(1)}(0) \gamma^{i}_{\da a}
\theta^{(1)}_{a}(0))
\langle \tilde{w}(C_1) \cdots \tilde{w}(C_l) \rangle\n
&&-\frac{1}{I} \int_{0}^{p^{(1)+}} d\sigma \lambda^{(1)}_{\da}(\sigma)
\big \langle
\tilde{\rm Tr}\big(P e^{i\int_{0}^{\sigma}d\sigma'
(k^{(1)\mu}(\sigma')A_{\mu}+\bar{\lambda}^{(1)}(\sigma')\psi)}\big)\nonumber\\
&&\times \tilde{\rm Tr}\big(P e^{i\int_{\sigma}^{p^{(1)+}}d\sigma''
(k^{(1)\mu}(\sigma'')A_{\mu}+\bar{\lambda}^{(1)}(\sigma'')\psi)}\big)
\tilde{w}(C_2) \cdots \tilde{w}(C_l)\big \rangle \n
&&- \frac{1}{I}\sum_{b=2}^{l}
\int_{0}^{p^{(b)+}} d\sigma \lambda^{(b)}_{\da}(\sigma)
\big\langle
\tilde{\rm Tr}\big(P e^{i\int_{\sigma}^{p^{(b)+}}d\sigma'
(k^{(b)\mu}(\sigma')A_{\mu}+\bar{\lambda}^{(b)}(\sigma')\psi)}
v(C_1)\nonumber\\
&&\times P e^{i\int_{0}^{\sigma}d\sigma''
(k^{(b)\mu}(\sigma'')A_{\mu}+\bar{\lambda}^{(b)}(\sigma'')\psi)}\big)
\tilde{w}(C_2) \cdots \check{\tilde{w}}(C_b)  \cdots \tilde{w}(C_l) \big\rangle,
\label{eqlambda}
\eeqa
\beqa
&&\hspace*{-1cm}\theta^{(1)}_{\dot{a}}(0)
\langle \tilde{w}(C_1) \cdots \tilde{w}(C_l) \rangle\n
&=& \frac{g^2 I}{\epsilon} (x^{'(1)i}(0) \gamma^{i}_{a \da})^{-1}
\bigg\{ \lambda^{(1)}_{a}(0) \langle \tilde{w}(C_1) \cdots \tilde{w}(C_l) \rangle\n
&&+\frac{1}{I} \int_{0}^{p^{(1)+}} d\sigma \lambda^{(1)}_{a}(\sigma)
\big\langle
\tilde{\rm Tr}\big(P e^{i\int_{0}^{\sigma}d\sigma'
(k^{(1)\mu}(\sigma')A_{\mu}+\bar{\lambda}^{(1)}(\sigma')\psi)}\big)\nonumber\\
&&\times \tilde{\rm Tr}\big(P e^{i\int_{\sigma}^{p^{(1)+}}d\sigma''
(k^{(1)\mu}(\sigma'')A_{\mu}+\bar{\lambda}^{(1)}(\sigma'')\psi)}\big)
\tilde{w}(C_2) \cdots \tilde{w}(C_l)\big \rangle \n
&&+ \frac{1}{I}\sum_{b=2}^{l}
\int_{0}^{p^{(b)+}} d\sigma \lambda^{(b)}_{a}(\sigma)
\big\langle
\tilde{\rm Tr}\big(P e^{i\int_{\sigma}^{p^{(b)+}}d\sigma'
(k^{(b)\mu}(\sigma')A_{\mu}+\bar{\lambda}^{(b)}(\sigma')\psi)}
v(C_1)\nonumber\\
&&\times P e^{i\int_{0}^{\sigma}d\sigma''
(k^{(b)\mu}(\sigma'')A_{\mu}+\bar{\lambda}^{(b)}(\sigma'')\psi)}\big)
\tilde{w}(C_2) \cdots \check{\tilde{w}}(C_b)  \cdots \tilde{w}(C_l)\big\rangle \bigg\}.
\label{eqtheta}
\eeqa
We eliminate a half of fermionic degrees of freedom
by using the above two equations just like
eliminating the half of the fermionic degrees of freedom
in the light-cone
field theory by using the equation of motion.
We can  also rewrite Eq. (\ref{localrepinv}) as
\beqa
&&x^{'(1)-}(\sigma)\langle \tilde{w}(C_1) \cdots \tilde{w}(C_l) \rangle\n
&&=(k^{(1)i}(\sigma)x^{'(1)i}(\sigma)
+i \lambda^{(1)}_{a}(\sigma) \theta^{(1)}_{a}(\sigma)
 +i \lambda^{(1)}_{\da}(\sigma) \theta^{(1)}_{\da}(\sigma))
\langle \tilde{w}(C_1) \cdots \tilde{w}(C_l) \rangle.\nonumber\\
\label{lightconerepinv}
\eeqa

Our task is summarized as follows. By using Eqs.~(\ref{eqk-}), (\ref{eqlambda}),
(\ref{eqtheta}) and (\ref{lightconerepinv}) iteratively and repeatedly, we
eliminate $\lambda_{n\da}$,
$\theta_{n\da}$ and $x^{'-}$,
and evaluate the light-cone Hamiltonian,
\beq
H=\sum_{b}\int d\sigma k^{(b)-}(\sigma).
\eeq
In this process, various interaction terms of
order $1/I^k \sim O(1/N^{ka})$ are generated, which represent processes where
the $k+2$ strings interact at one point, {i.e.,} $(k+2)$-Reggeon vertices.
In this procedure, we should take the continuum limit
by keeping the higher
order terms of $\ep$ vanishing in the naive $\ep \rightarrow 0$ limit.
This task is in general very hard to perform because we should treat
infinite series. However we will discuss in \S 2.4 that
the continuum limit is completely controllable by an analysis based on
a power counting of $\ep$ and the symmetries.
We should consider the supercharges first rather than the
Hamiltonian in this more rigorous treatment.
In the next subsection, we first sketch how we can derive
light-cone Hamiltonian for type IIB superstring from the loop equations.

\subsection{Light-cone Hamiltonian}
Here we neglect the interaction terms in Eq.~(\ref{eqk-}) and
concentrate on the free part of light-cone Hamiltonian.
We can see that if we ignore the non-quadratic terms and set
$\frac{g^2 I}{\ep} \sim \alpha^{'2} \sim$ const,
the free part of Eq.~(\ref{eqk-}) has the same form as
the free light-cone Hamiltonian of type IIB superstring
in the naive continuum limit except lacking the
$\lambda'_{a} \lambda_{a}$ term. In the following,
we verify that we indeed
obtain this term when we eliminate $\theta_{\da}$
in the free part of Eq.~(\ref{eqk-}) using Eq.~(\ref{eqtheta}).
Let us consider the naive leading contribution
in the free part of Eq.~(\ref{eqtheta}):
\beq
\th^{(1)}_{\da}(\sigma)\langle w(C_1) \cdots w(C_l)\rangle
={x^{'(1)}_{i}(\sigma) \gamma ^i_{\da a}\over x^{'(1)}_{i}
(\sigma)^2}
{g^2I\over \ep}\la^{(1)}_{a}(\sigma)
\langle w(C^1)w(C^2) \cdots w(C^l) \rangle.
\label{eqtha}
\eeq
We assume first that the free part of Eq.~(\ref{eqk-})
correctly describes the free part of light-cone Hamiltonian.
This assumption can be justified by showing that the non-quadratic
terms are indeed negligible in the continuum limit except for finite
renormalization of the quadratic terms.
Since we are dealing with free two-dimensional field theories,
we can use standard techniques of conformal field theory
to estimate the effects of the non-quadratic terms.
We note that the $x^{'(1)}_{i}(\sigma)^2$ is of order $1/\ep^2$ since
we have a cutoff length $\ep$.
So we may expand
$x^{'(1)}_{i}(\sigma)^2 = 1/ \alpha \ep^2 +:x^{'(1)}_{i}(\sigma)^2 :$,
where $:y:$ denotes the normal ordered operator constructed out of $y$.
The symbol $\alpha$ in the denominator is a quantity proportional to $\sqrt{\ep/g^2 I}$
on dimensional grounds.
In this way, the left-hand side of Eq.~(\ref{eqtha}) becomes
\beq
\ep^2{\alpha g^2 I\over {\ep}}
x^{'(1)}_{i} (\sigma )\gamma ^i_{\da a}(1-\ep^2\alpha
:x^{'(1)}_{i} (\sigma)^2 :
+\cdots )\la^{(1)}_{a}(\sigma)
\langle w(C_1) \cdots w(C_l) \rangle.
\eeq
For example, a typical non-quadratic term
in the free part of Eq.~(\ref{eqk-}),
\beq
\ep^4k^{(1)i}(\sigma)^2 x^{'(1)j}(\sigma)\gamma^{j}_{\da a}
\lambda^{'(1)}_{a}(\sigma) x^{'(1)k}(\sigma)\gamma^{k}_{\da b}
\lambda^{'(1)}_{b}(\sigma),
\eeq
generates the $\lambda^{'(1)}_{a}(\sigma)\lambda^{(1)}_{a}(\sigma)$ term
due to the order $1/\ep^4$ divergence from the operator product
expansion of
$k^{(1)i}(\sigma)^2 x^{'j(1)}(\sigma) x^{'(1)k}(\sigma)$.

In this way, we see that non-quadratic higher-dimensional operators do
not appear in the continuum limit due to the suppression of the powers of $\ep$
and only renormalize finitely the quadratic operators. Therefore we expect
to obtain the following  free Hamiltonian in the
continuum limit:
\beqa
H_{\rm free}&=& \sum_{b}\int _0^{p^{(b)+}} d\sigma {1\over 2}
\bigg(c k^{(b)i}(\sigma)^2 +c'\frac{\ep}{g^2 I} x^{'(1)i} (\sigma)^2
-ic''\lambda^{'(b)}_{a}(\sigma)\lambda^{(b)}_{a}(\sigma)\nonumber\\
&&\hspace{3cm}-ic'''\frac{1}{g^2 N}
\theta^{'(b)}_{a}(\sigma)\theta^{(b)}_{a}
(\sigma)\bigg).
\label{freeHam}
\eeqa
Here we assume that terms with the negative powers of $\ep$ and other finite
terms such as $k^i x'^i$ do not remain, which is guaranteed by an argument
based
on a power counting and symmetries as is seen in the next subsection.
In the bosonic model, this is not guaranteed and the $1/\ep^2$ divergence
remains in general. This is the reason why we cannot obtain the light-cone
string field theory from the bosonic reduced model.
The Hamiltonian (\ref{freeHam})
is identical to that of type IIB superstring theory\cite{GSB}
if $\lambda_{a}(\sigma )$ and $\th _{a}(\sigma )$ are rescaled
appropriately and rotated
by a complex phase factor $\eta = \exp({i\pi\over 4})$ as follows:
\beq
\la _{a}(\sigma ) \rightarrow \eta \la _{a}(\sigma ) ,
~\th _{a}(\sigma ) \rightarrow \eta^*  \th _{a}(\sigma ) .
\label{phase}
\eeq
We elaborate more on this point in connection with the supercharges
in the next subsection.
Here we cannot determine the coefficients $c$, $c'$, $c''$ and $c'''$, which
include the effects of renormalizations, and in the next subsection we can
fix them by using $\mathcal{N}$=2 supersymmetry.
From Eq.~(\ref{freeHam}) we find $\alpha '^2 \sim g^2 I/\ep$ and hence
we should obtain the prescription of the scaling limit,
$g^2 N^{a+b} \sim \alpha '^2 \sim$ const.

\subsection{General proof}
In this subsection, we give a general proof of our assertion that the
light-cone string field theory for type IIB superstring can be derived
from the loop equations of the IIB matrix model. We use a power counting and
a symmetry analysis
based on $\mathcal{N}$=2
supersymmetry, $SO(8)$ invariance and the parity symmetry on the string
worldsheet.

\subsubsection{Power counting and parity symmetry}
In order to perform a power counting for $\ep$, we first introduce
a mass dimension on the worldsheet through
the relation $[\ep]=-1$ and determine the dimension of each field.
The IIB matrix model action (\ref{action}) is decomposed into
\beqa
S & = &
-{1\over g^2} {\rm Tr}\bigg(-\frac{1}{2}[A^+,A^-]^2
-[A^+,A^i][A^-,A^i]+{1\over 4}[A^i,A^j]^2 \n
& & +{1\over 2}(\sqrt{2}\psi _{{a}}[A^+ ,\psi _{{a}}]
-\psi_{a}\gamma ^i_{{a}\da}[A^i,\psi _{\da}]
-\psi_{\da}\gamma ^i_{\da{a}}[A^{i},\psi _{{a}}]
+\sqrt{2}\psi _{\da}[A^- ,\psi _{\da}])\bigg ).\nonumber\\
\label{SO(8)decomposition}
\eeqa
By demanding the IIB matrix model action (\ref{SO(8)decomposition})
to be dimensionless, we obtain
\beq
[A^i]=0, \: [A^+]=-[A^-], \: [A^+]=-2[\psi_a] \quad \mbox{and} \quad
                             [A^-]=-2[\psi_{\da}].
\eeq
On the other hand, the Wilson loop (\ref{Wilsonloop}) is decomposed
into
\beqa
w(C)&=&{\rm Tr}\big(P \exp\big ( i\int_{0}^{p^+}d\sigma (-k^+(\sigma)A^- 
\nonumber\\
&&-k^-(\sigma) A^+
+k^i(\sigma) A^i -i \lambda_{a}(\sigma) \psi_{a}
-i \lambda_{\da}(\sigma) \psi_{\da}\big)\big).
\eeqa
From this, we also
read off the relations
\beqa
& &[k^{+}]+[A^-]=1, \n
& &[k^{-}]+[A^+]=1, \n
& &[k^{i}]+[A^i]=1, \n
& &[\lambda_{a}]+[\psi_a]=1, \n
& &[\lambda_{\da}]+[\psi_{\da}]=1.
\eeqa
Noting that we should set $[k^{+}]$ to be zero since $k^{+}(\sigma)=1$
in our light-cone setting, we can determine
the dimensions of all quantities as follows:
\beqa
& &[k^{+}]=0,\: [k^{-}]=2,\: [k^{i}]=1, \:
[\lambda_{a}]=\frac{1}{2}, \: [\lambda_{\da}]=\frac{3}{2}, \n
& &[x^{+}]=-1,\: [x^{-}]=1,\: [x^{i}]=0, \:
[\theta_{a}]=\frac{1}{2}, \: [\theta_{\da}]=-\frac{1}{2}, \n
& &[A^+]=-1, \: [A^-]=1, \: [A^i]=0, \: [\psi_a]=\frac{1}{2},
\: [\psi_{\da}]=-\frac{1}{2}.
\label{dimension}
\eeqa

Next we define a symmetry which corresponds to the parity
on the string worldsheet. It is seen easily
that the IIB matrix model action (\ref{action}) is formally invariant
under the following
transformation:
\beqa
& &A_{\mu} \rightarrow A_{\mu}^{t}, \n
& &\psi \rightarrow -i\psi^t.
\label{paritytransfmatrix}
\eeqa
This transformation flips the direction of the Wilson loop in the
following way:
\beqa
w(C)&=&{\rm Tr}\left(Pe^{i\int_{0}^{p^+}d\sigma
(k^{\mu}(\sigma)A_{\mu}+\bar{\lambda}(\sigma) \psi)}\right)\nonumber\\
&&\hspace{2cm}\rightarrow
{\rm Tr}\left(P\prod^{M}_{n=1}e^{i\int_{0}^{p^+}d\sigma(k^{\mu}(p^+-\sigma)A_{\mu}
                            -i\bar{\lambda}(p^+-\sigma) \psi)}\right).
\eeqa
Therefore our theory has a symmetry under the transformation
\beqa
& &k^{\mu}(\sigma) \rightarrow k^{\mu}(p^+-\sigma), \n
& &\lambda(\sigma) \rightarrow i\lambda(p^+-\sigma),
\label{paritytransffield1}
\eeqa
which we identify with the worldsheet parity. We also obtain
the parity transformation for the dual variables $x^{\mu}_{n}$ and
$\theta_n$:
\beqa
& &x^{\mu}(\sigma) \rightarrow x^{\mu}(p^+-\sigma), \n
& &\theta(\sigma) \rightarrow -i\theta(p^+-\sigma).
\label{paritytransffield2}
\eeqa

\subsubsection{$\mathcal {N}$=2 supersymmetry}
We denote the supercharges generating the transformations
\beq
\cases{
\delta^{(1)} A_\mu &  $= i \bar{\varepsilon}\Gamma_\mu \psi $\cr
\delta^{(1)} \psi & $=\frac{i}{2}\Gamma^{\mu\nu} [A^\mu, A^\nu] {\varepsilon}$
}  \label{super1}
\eeq
and
\beq
\cases{
\delta^{(2)} A_\mu & = 0  \cr
\delta^{(2)} \psi & $=  \xi$,
}
\label{super2}
\eeq
by $Q^{(1)}$ and $Q^{(2)}$ respectively.
We can determine the dimensions and parities of the parameters $\varepsilon$
and $\xi$ in Eqs. (\ref{super1}) and (\ref{super2})
by comparing their both sides,
\beqa
& &[\varepsilon_{a}]=[\xi_{a}]=\frac{1}{2}, \:
[\varepsilon_{\da}]=[\xi_{\da}]=-\frac{1}{2}, \n
& &\varepsilon_a \rightarrow i\varepsilon_a, \:
   \varepsilon_{\da} \rightarrow i\varepsilon_{\da},\n
& &\xi_a \rightarrow -i\xi_a, \:
   \xi_{\da} \rightarrow -i\xi_{\da}.
\label{SUSYparameter}
\eeqa
This fixes the dimensions and parities of the supercharges $Q^{(1)}$
and $Q^{(2)}$
since $\varepsilon_a Q^{(1)}_{a}+\varepsilon_{\da} Q^{(1)}_{\da}
+\xi_a Q^{(2)}_{a}+\xi_{\da} Q^{(2)}_{\da}$ generates the transformations
(\ref{super1}) and (\ref{super2}):
\beq
[Q^{(1)}_{a}]=[Q^{(2)}_{a}]=-\frac{1}{2}, \:
[Q^{(1)}_{\da}]=[Q^{(2)}_{\da}]=\frac{1}{2},
\label{superchargedimension}
\eeq
\beqa
& &Q^{(1)}_{a} \rightarrow -iQ^{(1)}_{a}, \:
   Q^{(1)}_{\da} \rightarrow -iQ^{(1)}_{\da},\n
& &Q^{(2)}_{a} \rightarrow iQ^{(2)}_{a}, \:
   Q^{(2)}_{\da} \rightarrow iQ^{(2)}_{\da}.
\label{superchargeparity}
\eeqa
Here we note that Eqs. (\ref{superchargedimension}) and
(\ref{superchargeparity})
are consistent with the anti-commutation relations
\beqa
& &\{Q^{(1)},Q^{(1)}\}=0, \: \{Q^{(2)},Q^{(2)}\}=0, \n
& &\{Q^{(1)}_{a},Q^{(2)}_{b}\}=\sqrt{2} P^+\delta_{ab}, \n
& &\{Q^{(1)}_{a},Q^{(2)}_{\da}\}=P^i\gamma^{i}_{a\da}, \:
\{Q^{(1)}_{\da},Q^{(2)}_{a}\}=P^i\gamma^{i}_{\da a}, \n
& &\{Q^{(1)}_{\da},Q^{(2)}_{\dot{b}}\}=\sqrt{2} H\delta_{\da\dot{b}}.
\label{SUSYanticom}
\eeqa

\subsubsection{Free parts of supercharges and Hamiltonian}
The supercharges $Q^{(1)}$ and $Q^{(2)}$ can be expressed as
differential operators
on the loop space using the Ward identities. In principle we can
eliminate the operators
$k^{-}$, $x^{'-}$,
$\lambda_{\da}$ and $\theta_{\da}$ by repeatedly using
the loop equations and the reparametrization invariance as is discussed for
$k^{-}$ in the previous section. Note that we obtain interaction terms
through this procedure.
However as we will see just below, the
forms of their continuum limit are completely determined by the dimension,
parity and $SO(8)$ invariance.
First we concentrate on free parts of the supercharges $Q^{(1)}$ and $Q^{(2)}$,
{i.e.},
ignore the interaction terms.
By using the power counting, Eqs. (\ref{dimension}) and
(\ref{superchargedimension}),
$SO(8)$ invariance and the parity symmetry, Eqs. (\ref{paritytransffield1}),
(\ref{paritytransffield2}) and
(\ref{superchargeparity}), we can deduce the following
forms of free supercharges in the $\ep \rightarrow 0$ limit:
\beqa
& &Q^{(1)}_{{\rm free} \, a}=\int d\sigma a_1 \theta_a(\sigma), \n
& &Q^{(2)}_{{\rm free} \, a}=\int d\sigma a_2 \lambda_a(\sigma), \n
& &Q^{(1)}_{{\rm free} \, \da}=\int d\sigma
                       (b_1x'^i(\sigma)\gamma^{i}_{\da a}\lambda_a(\sigma)
                       +c_1k^i(\sigma)\gamma^{i}_{\da a}\theta_a(\sigma)), \n
& &Q^{(2)}_{{\rm free} \, \da}=\int d\sigma
                       (b_2k^i(\sigma)\gamma^{i}_{\da a}\lambda_a(\sigma)
                       +c_2x'^i(\sigma)\gamma^{i}_{\da a}\theta_a(\sigma)).
\label{freesupercharge}
\eeqa
In Eq. (\ref{freesupercharge}) we have excluded
terms such as $\frac{1}{\ep}x^i\gamma^i\lambda$ by translation invariance.
It is easy to see that all possible terms which appear with negative powers
of $\ep$ are forbidden by the symmetries. In this sense the existence of the
continuum limit is guaranteed by the symmetries. We can also fix
undetermined coefficients in Eq. (\ref{freesupercharge})
by the $\mathcal{N}$=2 supersymmetry (\ref{SUSYanticom}) as follows.
From $\{Q^{(1)}_{a},Q^{(2)}_{b}\}=\sqrt{2}P^+\delta_{ab}$,
$\{Q^{(1)}_{a},Q^{(2)}_{\da}\}=P^i\gamma^{i}_{a\da}$ and
$\{Q^{(1)}_{\da},Q^{(2)}_{a}\}=P^i\gamma^{i}_{\da a}$, we obtain
\beq
a_1a_2=\sqrt{2}, \: a_1b_2=1 \quad \mbox{and} \quad a_2c_1=1.
\eeq
Therefore Eq. (\ref{freesupercharge}) is reduced to
\beqa
& &Q^{(1)}_{{\rm free} \, a}=a_1\int d\sigma \theta_a(\sigma), \n
& &Q^{(2)}_{{\rm free} \, a}=\frac{\sqrt{2}}{a_1}\int d\sigma \lambda_a(\sigma), \n
& &Q^{(1)}_{{\rm free} \, \da}=\int d\sigma
               \left(\frac{a_1}{\sqrt{2}}k^i(\sigma)\gamma^{i}_{\da a}
\theta_a(\sigma)
                  +b_1x'^i(\sigma)\gamma^{i}_{\da a}\lambda_a(\sigma)\right), \n
& &Q^{(2)}_{{\rm free} \, \da}=\int d\sigma
             \left(\frac{1}{a_1}k^i(\sigma)\gamma^{i}_{\da a}\lambda_a(\sigma)
                       +c_2x'^i(\sigma)\gamma^{i}_{\da a}\theta_a(\sigma)\right).
\label{freesuperchargefinal}
\eeqa
The free part of the Hamiltonian
$\int d\sigma k^{-}(\sigma)$ is obtained by
$\{Q^{(1)}_{\da},Q^{(2)}_{\dot{b}}\}=\sqrt{2}H\delta_{\da\dot{b}}$ as
\beq
H_{\rm free}=\int d\sigma \left(\frac{1}{2}k^i(\sigma)^2+\frac{1}{\sqrt{2}}
b_1c_2x'^i(\sigma)^2
-i\frac{b_1}{\sqrt{2}a_1}\lambda'_a(\sigma)\lambda_a(\sigma)
           -i\frac{a_1c_2}{2}c_2\theta'_a(\sigma)\theta_a(\sigma)\right).\nonumber\\
\label{freehamiltonian}
\eeq
In order to compare these results with the Green-Schwarz light-cone formalism,
we redefine the fermionic variables as
\beq
\lambda_a = \sqrt{\frac{a_1}{b_1}} \eta \check{\lambda}_a \quad
\mbox{and} \quad \theta_a = \sqrt{\frac{b_1}{a_1}} \eta^* \check{\theta}_a,
\label{fermionrescaling}
\eeq
where $\eta=e^{\frac{\pi i}{4}}$.
We also introduce rescaled supercharges $\check{Q}^1$ and $\check{Q}^2$ by
\beq
Q^{(1)} = \sqrt{a_1b_1} \eta^{*} \check{Q}^{1} \quad \mbox{and} \quad
Q^{(2)} = \frac{\eta}{\sqrt{a_1b_1}} \check{Q}^{(2)}.
\label{superchargerescaling}
\eeq
In terms of these new quantities, Eqs.~(\ref{freesuperchargefinal})
and (\ref{freehamiltonian}) become
\beqa
& &\check{Q}^{1}_{{\rm free} \, a}=\int d\sigma \check{\theta}_a(\sigma), \n
& &\check{Q}^{(2)}_{{\rm free} \, a}=\sqrt{2}\int d\sigma
\check{\lambda}_a(\sigma), \n
& &\check{Q}^{1}_{{\rm free} \, \da}=\int d\sigma
        \left(\frac{1}{\sqrt{2}}k^i(\sigma)\gamma^{i}_{\da a}
\check{\theta}_a(\sigma)
                  +ix'^i(\sigma)\gamma^{i}_{\da a}\check{\lambda}_a(\sigma)\right),
\n
& &\check{Q}^{(2)}_{{\rm free} \, \da}=\int d\sigma
             (k^i(\sigma)\gamma^{i}_{\da a}\check{\lambda}_a(\sigma)
              -ib_1c_2x'^i(\sigma)\gamma^{i}_{\da a}\check{\theta}_a(\sigma)),
\n
& &H_{\rm free}=\int d\sigma \left(\frac{1}{2}k^i(\sigma)^2
                           +\frac{1}{\sqrt{2}}b_1c_2x'^i(\sigma)^2
+\frac{1}{\sqrt{2}}\check{\lambda}'_a(\sigma)\check{\lambda}_a(\sigma)
    -\frac{1}{2}b_1c_2\check{\theta}'_a(\sigma)\check{\theta}_a(\sigma)\right),
\nonumber\\
\eeqa
which completely agree with the light-cone Green-Schwarz free Hamiltonian
and supercharges for type IIB superstring. This fact also justifies
the analytic
continuation introduced for fermionic fields in Ref. \citen{IKKT}.
We also note that we have obtained the relation $b_1c_2 \sim 1/\alpha'^2$,
and $b_1c_2$ should be equal to $\ep/g^2 I$ multiplied by some numerical
constant
as is illustrated in the previous subsection.

\subsubsection{Interaction parts of supercharges and Hamiltonian}
In this subsection, we examine the structure of the interaction parts of
the supercharges and the Hamiltonian. First we consider the contributions
of order $1/I$, which correspond to 3-Reggeon vertices in string
field theory.
Since our free Hamiltonian is equal to that of the Green-Schwarz light-cone
formalism and the interactions of loops are local in
our loop equations,
we can use the same arguments as in light-cone string field theory.
In general, the operators inserted near the interaction points in 3-Reggeon
vertices generate divergences coming from the Mandelstam mapping.
Since our Wilson loops are written by the variables $k^i$ and $\lambda$,
the corresponding 3-Reggeon vertices should consist of delta functions
representing the matching of three strings in the $k-\lambda$ space,
which is the same as in Ref. \citen{GSB}.
Therefore the $k^i$, $x'^i$ and $\lambda_a$ diverge as $1/\sqrt{\ep}$ near
the interaction points while $\theta_a$ is of order $\ep^0$ there.
We also note that every derivative of $\sigma$ acting on the fields introduces
an extra factor of $1/\sqrt{\ep}$.
Therefore the interaction part at order $1/I$ of the supercharges possesses
the following general structure:
\beqa
&&\frac{1}{I}\int d\sigma \int d\sigma_1 \ep^{\eta}(k^i)^{\alpha} (x'^i)^{\beta}
(\lambda_a)^{\gamma}(\theta_a)^{\delta}\nonumber\\
&&\hspace*{1cm}\times(\mbox{derivative})^{\kappa}
(\mbox{products of delta functions for }k^i \mbox{ and } \lambda_a),
\eeqa
where $k^i$, $x'^i$, $\lambda_a$ and $\theta_a$ represent the operators
inserted
near the interaction points, and $\kappa$ is the total number of derivatives
acting on these operators.

For example, let us consider the interaction part of $Q^{(1)}_{\da}$.
In this case, the dimensional analysis
$[Q^{(1)}_{{\rm int} \, \da}]=[Q^{(1)}_{{\rm free} \,\da}]$ leads to
$-\eta+\alpha+\beta+\frac{1}{2}\gamma+\frac{1}{2}\delta
+\kappa-1=\frac{1}{2}$,
where we used the relation $[I]=-1$ and the fact that the dimensions
of the delta functions are canceled due to supersymmetry.
Therefore
the total powers of $\ep$ which appear in the interaction part of
 $Q^{(1)}_{\da}$
is evaluated as
\beqa
\zeta&=&\eta-\frac{1}{2}\alpha-\frac{1}{2}\beta-\frac{1}{2}\gamma
                                -\frac{1}{2}\kappa \n
     &=&\frac{1}{2}\alpha+\frac{1}{2}\beta+\frac{1}{2}\delta+\frac{1}{2}\kappa
                                -\frac{3}{2}.
\eeqa
The case in which $\alpha=\beta=\delta=\kappa=0$ is excluded by
$SO(8)$ invariance. We can consider four cases in which $\zeta=-1$:
(1)$\alpha=1$ and $\beta=\delta=\kappa=0$, (2)$\beta=1$ and
$\alpha=\delta=\kappa=0$, (3)$\delta=1$ and $\alpha=\beta=\kappa=0$ and
(4)$\kappa=1$ and $\alpha=\beta=\delta=0$.  Cases (3) and (4) are not
permitted by $SO(8)$ invariance. If we take the large-$N$ limit with $I \ep$
kept fixed, Cases (1) and (2) survive in the $\ep \rightarrow 0$ limit.
This limit corresponds to taking $g_{\rm st} \sim 1/I \ep$.
Note that in this limit all of the other cases vanish because
$\zeta$ is larger
than $-1$ for them.
Furthermore we can restrict the values of
$\gamma$ by the parity symmetry and deduce the structure
of $Q^{(1)}_{{\rm int} \,\da}$ as follows:
\beqa
& &Q^{(1)}_{{\rm int} \,\da}=\frac{1}{I \ep}\int d\sigma \int d\sigma_1
(\sqrt{\ep}k^i((\sqrt{\ep}\lambda_a)^3+(\sqrt{\ep}\lambda_a)^7)
+\sqrt{\ep}x'^i(\sqrt{\ep}\lambda_a+(\sqrt{\ep}\lambda_a)^5)) \n
& & \hspace{3cm}\times
(\mbox{products of delta functions for }k^i \mbox{ and } \lambda_a).
\label{Q1intda}
\eeqa
This structure agrees with that of the light-cone
string field theory.\cite{GSB} Applying a similar analysis to
$Q^{(2)}_{{\rm int} \,\da}$, we obtain
\beqa
& &Q^{(2)}_{{\rm int} \,\da}=\frac{1}{I\ep}\int d\sigma \int d\sigma
(\sqrt{\ep}k^i(\sqrt{\ep}\lambda_a+(\sqrt{\ep}\lambda_a)^5)
+\sqrt{\ep}x'^i((\sqrt{\ep}\lambda_a)^3+(\sqrt{\ep}\lambda_a)^7)) \n
& & \hspace{3cm}\times
(\mbox{products of delta functions for }k^i \mbox{ and } \lambda_a),
\label{Q2intda}
\eeqa
which also agrees with the light-cone string field theory.
As for $Q^{(1)}_{a}$ and $Q^{(2)}_{a}$, no $1/I$ contribution remains non-zero
in this limit since the minimum value of $\zeta$ is $-\frac{1}{2}$ in these
cases. Therefore we conclude that
$Q^{(1)}_{{\rm int} \,a}$ and $Q^{(2)}_{{\rm int} \,a}$ are equal to zero at order $1/I$,
which is again consistent
with the light-cone string field theory.
Note that the right-hand sides of Eqs. (\ref{Q1intda}) and (\ref{Q2intda})
are uniquely determined
by $\mathcal{N}$=2 supersymmetry, as is shown in Ref. \citen{GSB}.
Finally the anti-commutation relation
$\{Q^{(1)}_{\da},Q^{(2)}_{\dot{b}}\}=2H\delta_{\da\dot{b}}$ fixes
the interaction part of $H$, which is certainly consistent with
the light-cone string field theory.

Next we consider the contributions of order $1/I^k \: (k\geq2)$,
which correspond to $(k+2)$-Reggeon vertices. The general
structure of the interaction part is represented as
\beqa
& &\frac{1}{I^k}\int d\sigma \int d\sigma_1 \cdot \cdot \cdot d\sigma_{k}
       \ep^{\eta}(k^i)^{\alpha} (x'^i)^{\beta}
       (\lambda_a)^{\gamma}(\theta_a)^{\delta}(\mbox{derivative})^{\kappa} \n
& & \hspace{3cm}\times
(\mbox{products of delta functions for }k^i \mbox{ and } \lambda_a).
\eeqa
From the Mandelstam mapping in these cases, it is natural to consider that
the $k^i$, $x'^i$ and $\lambda_a$ diverge as $\ep^{-\frac{k}{k+1}}$ near
the interaction points.
Therefore the total power $\zeta$ of $\ep$ is evaluated as
\beq
\zeta=\frac{1}{k+1}(\alpha+\beta+\kappa)+\frac{1-k}{2(k+1)}\gamma
+\frac{1}{2}\delta-\rho,
\eeq
where $\rho$ is $\frac{1}{2}$ for $Q^{(1)}_{{\rm int} \,a}$ and
$Q^{(2)}_{{\rm int} \,a}$,
and $\frac{3}{2}$ for $Q^{(1)}_{{\rm int} \,\da}$ and $Q^{(2)}_{{\rm int} \,\da}$ and
the terms in which $\zeta\leq-k$ survive in the $\ep \rightarrow 0$ limit
if $I \ep$ is fixed.
It is verified easily that there are no surviving terms for any values of $k$
in $Q^{(1)}_{{\rm int} \,a}$ and $Q^{(2)}_{{\rm int} \,a}$
in the $\ep \rightarrow 0$ limit, which is consistent with the
light-cone string
field theory. Using $SO(8)$ invariance, we can show that
in $Q^{(1)}_{{\rm int} \,\da}$ and $Q^{(2)}_{{\rm int} \,\da}$ some terms with
$\gamma$ equal to
five might survive for $k=2$ and ones with $\gamma$ equal to seven
for $k=2$ and
$k=3$. Presumably it is not possible to satisfy
$\mathcal{N}$=2 supersymmetry only by these restricted terms.
Therefore we may conclude that there are no
contributions of order $1/I^k \: (k\geq2)$
in $Q^{(1)}_{{\rm int} \,\da}$, $Q^{(2)}_{{\rm int} \,\da}$ and the Hamiltonian,
which is also consistent with
the light-cone string field theory.

In this way, we almost confirm that our IIB matrix model reproduces
the light-cone string field theory for type IIB superstring.

We have the following two relations about the scaling limit.
\beq
\alpha^{'2} \sim \frac{g^2 I}{\ep},
\label{alphaprime}
\eeq
\beq
g_{\rm st} \sim \frac{1}{I \ep}.
\label{gst}
\eeq
Since we assume that
$I \sim N^a$ and $\ep \sim N^{-b}$,
we obtain from Eq.~(\ref{alphaprime}) the prescription of
the scaling limit:
\beq
g^2 N^{a+b} \sim {\rm const}.
\eeq
From Eq.~(\ref{gst}), in order to
obtain a finite string coupling,
we also have a relation
\beq
a=b.
\label{a=b}
\eeq
We need to estimate large $N$ behavior of $I$ and $\ep$,
which should be consistent with Eq.~(\ref{a=b}),
in order to fix the prescription of the scaling limit
completely.
Assuming, for example, that $\ep \sim \sqrt{\alpha'}/R$ and
$R \sim \sqrt{g} N^{\frac{1}{4}}$, we obtain as
a candidate
\beq
g^2 N^{\frac{1}{3}} \sim {\rm const}.
\eeq
\section{Dynamics of eigenvalues and space-time generation}
\setcounter{equation}{0}
In this section we analyze the structure of space-time,
and in particular, try to explain why our space-time is four-dimensional.
\cite{AIKKT}
As we mentioned in the introduction,
the diagonal elements of the bosonic matrices $A_\mu$
can be interpreted as space-time itself.
For example, if the diagonal elements distribute within a manifold
which extends in four dimensions but shrinks in six dimensions,
then a natural interpretation is that the space-time is four-dimensional.
We thus derive an effective theory for the diagonal elements,
and analyze their distribution.

We decompose $A_{\mu}$ into diagonal part $X_{\mu}$
and off-diagonal part $\tilde{A}_{\mu}$.
We also decompose $\psi$ into diagonal part $\xi$ and off-diagonal part $\tilde{\psi}$:
\beqa
A_{\mu}&=& X_{\mu} + \tilde{A}_{\mu}; \;\;
X_{\mu}= \left( \begin{array}{llll}
                 x_{\mu}^1 &&& \\
                 &x_{\mu}^2 && \\
                 && \ddots &   \\
                  &&& x_{\mu}^N
         \end{array} \right), \n
\psi &=& \xi + \tilde{\psi}; \;\;
\xi = \left( \begin{array}{llll}
                 \xi^1 &&& \\
                 &\xi^2 && \\
                 && \ddots &   \\
                  &&& \xi^N
         \end{array} \right),
\eeqa
where $x^i_\mu$ and $\xi^i_\alpha$ satisfy the constraints
$\sum_{i=1}^N x^i_\mu=0$ and $\sum_{i=1}^N \xi^i_\alpha =0$, respectively,
since we may fix the $U(1)$ part by translation invariance.
We then integrate out the off-diagonal parts $\tilde{ A_{\mu}}$
and $\tilde{\psi}$
and obtain the
effective action for supercoordinates of space-time $S_{\rm eff}[X,\xi]$.
The effective action for the space-time coordinates $S_{\rm eff}[X]$
can be obtained by further integrating out $\xi$:
\beqa
\int dAd\psi e^{-S[A,\psi]}
&=& \int dXd\xi e^{-S_{\rm eff}[X,\xi]} \nonumber \\
&=& \int dX e^{-S_{\rm eff}[X]},
\eeqa
where $dX$ and $d\xi$ stand for
$\prod_{i=1}^{N-1}\prod_{\mu=0}^{9} dx^i_\mu$
and $\prod_{i=1}^{N-1}\prod_{\alpha=1}^{16} d\xi^i_\alpha$, respectively.

We perform integrations over off-diagonal parts $\tilde{ A_{\mu}}$
and $\tilde{\psi}$ by the
perturbative  expansion in $g^2$,
which is valid
when all of the diagonal elements are widely separated from one another:
$|x^i -x^j| \gg g^{1/2}$.
After adding a gauge fixing and the Faddeev-Popov ghost term
\beq
S_{\rm g.f.} + S_{\rm F.P.} =
 -\frac{1}{2g^2} {\rm Tr} ([X_\mu, A^\mu]^2)
 - \frac{1}{g^2}{\rm Tr} ([X_\mu,b][A^\mu,c]),
 \eeq
the action  can be expanded   up to the second order of the
off-diagonal elements $\tilde{A_{\mu}}, \tilde{\psi}$ as
\begin{eqnarray}
S_2+S_{\rm g.f.}
&=&\frac{1}{2g^2} \sum_{i\ne j}
\bigl( (x_\nu^i-x_\nu^j)^2 {{\tilde A}^{ij}_\mu}{}^*
{{\tilde A}^{ij}}{}^\mu
- \bar{\tilde \psi}^{ji}\Gamma^\mu (x^i_\mu-x^j_\mu) \tilde \psi^{ij}\n
&&+ (\bar{\xi}^i-\bar{\xi}^j) \Gamma^\mu \tilde \psi^{ij}
{\tilde{A_\mu^{ij}}}^*
+\bar{\tilde \psi}^{ji} \Gamma^\mu ({\xi}^i-{\xi}^j){\tilde{A_\mu^{ij}}}
\bigr).
\label{eq:S2comp}
\end{eqnarray}
The first and the second terms are the kinetic
terms for $\tilde{A}$ and $\tilde{\psi}$ respectively,
while the last two terms are
$\tilde{A}\tilde{\psi}\xi$ vertices.
A  bosonic off-diagonal element $\tilde{A}_{\mu}^{ij}$
is transmuted to a fermionic
off-diagonal element $\tilde{\psi}^{ij}$
emitting a fermion zeromode $\xi^i$ or $\xi^j$.
This vertex conserves
indices for space-time points, $i$, $j$.
Note that the propagators for $\tilde{A}_\mu$ and $\tilde{\psi}$
behave as $\eta_{\mu\nu} /(x^i-x^j)^2$ and $(x^i_\mu -x^j_\mu)\Gamma^\mu
/(x^i-x^j)^2$
respectively,
thus they decrease in the long distances.

We obtain the
effective action for the zeromodes, $x_{\mu}^i$ and $\xi^i$,
at one-loop level:
\begin{eqnarray}
 \int d \tilde A d\tilde \psi d b dc \;
e^{ - (S_2 +S_{\rm g.f.}+S_{\rm F.P.})}
&=& \prod_{i<j} {\rm det}_{\mu\nu}
\bigl( \eta^{\mu\nu}+S^{\mu\nu}_{(ij)}\bigr)^{-1} \nonumber\\
&\equiv&
e^{-S_{\rm eff}^{\rm 1\mbox{-}loop} [X, \xi]} ,
\end{eqnarray}
where
\begin{equation}
S^{\mu\nu}_{(ij)}=\bar{\xi^{ij}} \Gamma^{\mu\alpha\nu}
\xi^{ij} \frac{x^{ij}_\alpha}{(x^{ij}) ^4}.  \label{eq:defS}
\end{equation}
Here $\xi^{ij}$ and $x_{\mu}^{ij}$ are abbreviations  for $\xi^i - \xi^j$
and $x_{\mu}^i- x_{\mu}^j $.
The effective action can be expanded as
\begin{eqnarray}
S_{\rm eff}^{\rm 1\mbox{-}loop} [X, \xi]
&=& \sum_{i<j} {\rm tr} \ln(\eta^{\mu\nu}+S^{\mu\nu}_{(ij)}) \nonumber \\
&=& - \sum_{i<j} {\rm tr}
\left(\frac{S^4_{(ij)}}{4}
+\frac{S^8_{(ij)}}{8} \right) ,
\end{eqnarray}
which is a sum of all pairs $(ij)$ of space-time points.
Here the symbol tr in the lower case stands for the trace over Lorentz
indices, $\mu$, $\nu$.
Other terms in the expansion vanish due to the properties of
Majorana-Weyl fermions in ten dimensions.

Note that the one-loop effective action
$S_{\rm eff}^{\rm 1\mbox{-}loop} [X, \xi]$ has ${\cal N}=2$ supersymmetry,
\beq
\cases{
\delta^{(1)}  x^i_\mu &  $= i \bar{\epsilon_1}\Gamma_\mu \xi^i $\cr
\delta^{(1)}  \xi ^i  &  $=0$
}, \,\,\,
\cases{
\delta^{(2)}  x^i_\mu &  $=0$ \cr
\delta^{(2)}  \xi ^i  &  $=\epsilon_2$
},
\label{eq:susya}
\eeq
which is a remnant of the one in the original theory,
\beq
\cases{
\delta^{(1)}  A_\mu &  $= i \bar{\epsilon_1}\Gamma_\mu \psi $\cr
\delta^{(1)}  \psi  &  $=\frac{i}{2}\Gamma^{\mu\nu}[A_\mu,A_\nu]\epsilon_1$
}, \,\,\,
\cases{
\delta^{(2)}  A_\mu &  $=0$ \cr
\delta^{(2)}  \psi  &  $=\epsilon_2$
}.
\eeq
Transformations for $\epsilon_1=\epsilon_2$ and
$\epsilon_1=-\epsilon_2$ correspond to those  generated by
${\cal N}=1$ supersymmetry generator $Q$ and its covariant derivative $D$.
In this sense zeromodes of  $x_i$ and $\xi_i$ may be viewed as
supercoordinates of ${\cal N}=1$
superspace.

The effective action for the space-time coordinates
$x^i_\mu$ is given by further integrating out
the fermion zeromodes  $\xi_i$:
\beqa
\int dX e^{-S_{\rm eff}^{\rm 1\mbox{-}loop}[X]}
&=&\int dXd\xi e^{-S^{\rm 1\mbox{-}loop}_{\rm eff}[X,\xi]} \n
&=& \int dXd\xi \prod_{i<j}\left[1+\frac{{\rm tr}(S_{(ij)}^4)}{4}
                             +\left(\frac{1}{2}\left(\frac{{\rm tr}(S_{(ij)}^4)}{4}\right)^2
                               +\frac{{\rm
tr}(S_{(ij)}^8)}{8}\right)\right].\nonumber\\
\label{eq:mulprod}
\eeqa
Here the products  are taken over all possible different
pairs $(ij)$ of the space-time points.
When we expand the multi-products, we select one of the three different
factors,
$1$, tr$(S^4_{(ij)})/4$  or  $({\rm tr}(S^8_{(ij)})/8+ ({\rm tr}(S^4_{(ij)}))^2/32 )$
for
each pair of $(ij)$. Since the last two  factors are functions of
$(x^i_\mu-x^j_\mu)$,
they can be visualized by bonds that  connect the space-time
points $x^i_\mu$ and $x^j_\mu$.
Since
the factors tr$(S^4_{(ij)})/4$
and  (tr$(S^8_{(ij)})/8+ ({\rm tr}(S^4_{(ij)}))^2/32 )$
contain 8 and 16 spinor components of $\xi^{ij}$,
we call them an 8-fold bond and a 16-fold bond, respectively.
We do not assign any  bond to the factor 1.
In this way  we can associate each term in the
expansion of multi-products in Eq. (\ref{eq:mulprod})  with a graph
connecting the space-time points by 8-fold bonds  and 16-fold  bonds.
Therefore the multi-products in Eq. (\ref{eq:mulprod}) can be replaced by a
summation over all possible graphs:
\beqa
\int dX e^{-S^{\rm 1\mbox{-}loop}_{\rm eff}[X]}
&=& \int dXd\xi \sum_{G:{\rm graph}}\;\; \prod_{(ij):{\rm bond\; of \;} G}\n
&&               \left[\left(\frac{{\rm tr}(S_{(ij)}^4)}{4}\right)\;\; {\rm or}\;\;
                \left(\frac{1}{2}\left(\frac{{\rm tr}(S_{(ij)}^4)}{4}\right)^2
                +\frac{{\rm tr}(S_{(ij)}^8)}{8}\right)\right].\nonumber\\
\label{eq:graph}
\eeqa
Here we sum over all possible graphs
consisting of 8-fold and 16-fold bonds.
For each bond $(ij)$ of $G$, we assign the first or the second factor
depending on whether it is an 8-fold or a 16-fold bond.

In order to saturate the grassmann integration $d\xi$,
we need $16(N-1)$ fermion zeromodes.
Since an 8-fold bond and a 16-fold bond contain 8 and 16 fermion zeromodes
respectively,
graphs remaining after the $\xi$ integrations are
those where the sum of the number of 8-fold bonds and twice the number of
16-fold bonds is equal to $2(N-1)$.
Thus, the number of bonds is of order $N$,
which is much
smaller than possible number of pairs $N(N-1)/2$.
In this sense $N$ space-time points are weakly bound.

Since 8-fold bond and 16-fold bond terms behave as
$(x^{ij})^{-12}$ and $(x^{ij})^{-24}$, both are strong attractive
interactions.
Hence, only the closer points can be connected by the bonds,
and these interactions become local.
On the other hand, since all possible graphs must be summed up,
this system has a permutation invariance among the points.
This reminds us of  summation over
all triangulations in the dynamical triangulation
approach to quantum gravity.
We come back to this analogy in \S 4.

In order to see some important features of the system (\ref{eq:graph}),
let us take an approximation.
If there were only 16-fold bonds,
considerable simplifications take place
and the $\xi$ integrations can be performed exactly.
Since a 16-fold bond term contains 16 fermion zeromodes,
it is proportional to delta function of the grassmann variables as
\beq
{\rm tr}(S_{(ij)}^8)
\sim \delta^{(16)}(\xi^i-\xi^j)\;\frac{1}{(x^i-x^j)^{24}}.
\eeq
If there is a loop in the graph, the contribution vanishes,
since the product of delta functions of grassmann variables on the loop
vanishes:
\beq
\delta^{(16)}(\xi^{i_1 i_2}) \delta^{(16)}(\xi^{i_2 i_3})\cdots
\delta^{(16)}(\xi^{i_k i_1}) = 0.
\eeq
Also, as we mentioned above, the remaining graphs have $(N-1)$ number of
16-fold bonds.
Hence, the remaining graphs are tree graphs which connect all $N$
points.
Such type of graphs are called ``maximal trees".
We also note that all maximal trees contribute equally
as we can see by performing $\xi$ integrations from the end points
of each maximal tree.
Therefore, only with 16-fold bonds,
the distribution of $x^i_\mu$ becomes ``branched polymer" type:
\beq
\int dX e^{-S^{\rm 16\mbox{-}fold\; bond}_{\rm eff}[X]}
= \int dX \sum_{G:{\rm maximal\; graph}}\;\;
              \prod_{(ij):{\rm bond\; of \;} G}
              \frac{1}{(x^i-x^j)^{24}}.
\label{eq:branced polymer}
\eeq

Note that all points are connected by the bonds, and
each $x^{ij}$ integration can be performed independently and
converges for large $x^{ij}$ on each bond.
Thus this system is infrared convergent.
As we proved rigorously in Ref. \citen{AIKKT},
this feature of IR convergence holds even with 8-fold bonds, and also,
to all orders in perturbation expansion.
This is consistent with the explicit calculations of the
partition function.\cite{MNS}\linebreak
This shows that all points
are gathered as a single bunch and hence space-time is inseparable.
Thus, the size and the dimensionality of the space-time can be
determined dynamically.
Note also that
the dynamics of branched polymer is well known
and its Hausdorff dimension is four.
It is conceivable that the smooth four-dimensional space-time
emerges by taking account of the effects from 8-fold bonds
as we argue in what follows.
Therefore, the model (\ref{eq:graph}) constitutes a candidate of
models for dynamical generation of four-dimensional space-time.

Before going into the analysis of the space-time structures
by using the effective action
(\ref{eq:graph}),
which is valid in the long distance,
we consider the short distance behavior of the system.
Let us suppose that a pair of the bosonic coordinates are degenerate
but the rest of the coordinates
are well separated from one another and from the center of mass
coordinates of the pair.
We can determine the
dynamics of the relative coordinates of the pair of the points,
from the exact solution for the $SU(2)$ case.
The distribution for the relative coordinates $r_{\mu}$ is
\beqa
&&\int d^{10}r f(r),\n
&& f(r) \sim \left\{ \begin{array}{ll}
                   1/r^{24} & r^2 \gg g \\
                   r^8      & r^2 \ll g
                \end{array}\right. .
\eeqa
We conclude that there is a pairwise repulsive potential
of $-8\:\ln r $ type
when two coordinates are close to each other.
It is clear that these considerations are valid
for arbitrary numbers of degenerate pairs although
the center of mass coordinates should be well separated.
Although it is possible to repeat these considerations
to the cases with higher degeneracy,
the analysis becomes more complicated.
Therefore we choose to adopt a phenomenological approach
and assume the existence of the hardcore repulsive potential
of the following form:
\beq
S_{\rm core}[X] = \sum_{i < j} g(x^i-x^j),
\label{eq:score}
\eeq
where
\beq
g(x^i-x^j)=\left\{ \begin{array}{ll}
                    -4 \ln ((x^i-x^j)^2/g) & {\rm for}\;(x^i-x^j)^2 \ll g \\
                    0                      & {\rm for}\;(x^i-x^j)^2 \gg g
                    \end{array}\right. .
\eeq
Hereafter, we investigate the structure of space-time
by using the one loop effective action (\ref{eq:graph})
plus the phenomenological hardcore potential (\ref{eq:score}).

If the number of 16-fold bonds is much larger than that of 8-fold bonds,
the system behaves as a branched polymer.
A small number of 8-fold bonds may fold the branched polymer into
a lower-dimensional manifold.
This can happen as in protein, a chain of amino acids is folded
into a lower-dimensional object like $\beta$-sheet,
by perturbative interactions.
Since the Hausdorff dimension of the branched polymer is four,
the core interactions exclude the manifolds in less than four dimensions.
Thus, four-dimensional space-time can be realized by this mechanism.
We are checking this conjecture by numerical simulations,
which we mention later in this section.

On the other hand, if the 8-fold bonds dominate,
the number of bonds are of order $2N$,
twice as much as in the above case.
Thus, the entropy of graph rearrangement becomes more important,
and the system might behave as a mean field phase,
where all the $N$ points condensate into a finite volume.
However, the core interactions prohibit an infinite density state,
and the system behaves as a droplet.
The 8-fold bond interaction can be written as
\begin{equation}
{\rm tr} (S_{(ij)}) ^4 \propto
 C^{\mu \nu \lambda \rho \alpha_1 ...\alpha_8}
\xi^{ij}_{\alpha_1} \cdots \xi^{ij}_{\alpha_8}
V^{ij}_{\mu \nu \lambda \rho}/(x^{ij})^{16},
\label{ST3}
\end{equation}
where $C^{\mu \nu \lambda \rho \alpha_1 ...\alpha_8}$ is an invariant
tensor, and $V^{ij}_{\mu \nu \lambda \rho}$ is a fourth rank
symmetric traceless
tensor constructed from $x^{ij}_\mu$.
Since $V^{ij}_{\mu \nu \lambda \rho}$ is traceless,
the average over orientations of $x^{ij}_\mu$ gives a suppression factor
for each 8-fold bond.
This suppression factor becomes weak if the system becomes 
lower-dimensional.
Also, the $\xi$ integrations give contractions
of $V^{ij}_{\mu \nu \lambda \rho}$ among different bonds,
which consist of inner products $x^{ij}_\mu x^{kl\; \mu}$
between different bonds.
These angle-dependent interactions may favor lower-dimensional space-time.

We can study self-consistently which phase is realized;
16-fold bond dominant branched polymer phase or 8-fold bond dominant
droplet phase.
The phase with the lowest free energy is realized.
However, in any phase, four-dimensional space-time can be realized
by one of the above mentioned mechanisms or by some combinations of them.
\begin{figure}[b]
\begin{center}
\leavevmode
\epsfxsize=\linewidth
\epsfbox{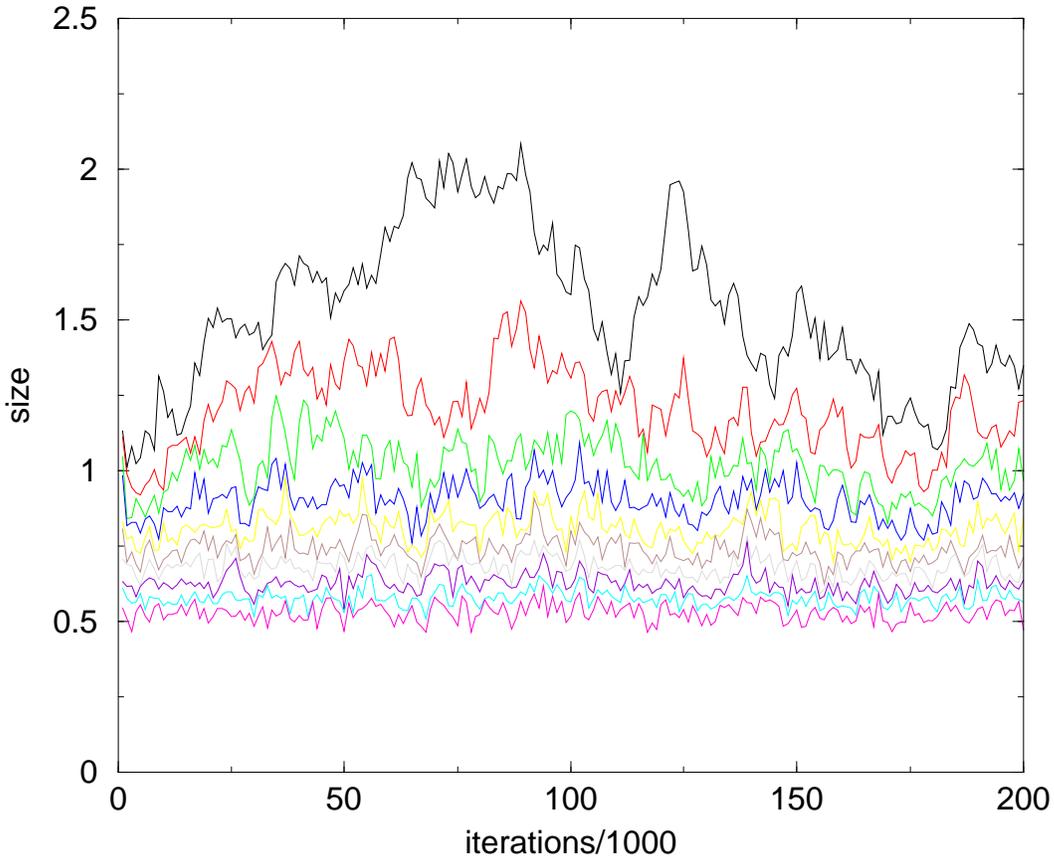}
\vspace{-0.3cm}
\caption{Result of the numerical simulation, where $N=800$
and the number of 8-fold bond is fixed to be 140.
The horizontal axis represents the number of iterations,
the vertical axis the length of the ten principal axes in ten dimensions.
Two of them are relatively large, suggesting the existence
of anisotropy in space-time.}
\label{fig:eigen}
\end{center}
\end{figure}

In the remainder of this section,
we show how we perform numerical simulations.
Our conjecture is that a small number of 8-fold bonds
fold the branched polymer into four-dimensional space-time.
Thus we take the following model:
\beq
Z=\int \prod_{i=1}^{N-1} d^{10}x^{i}
\sum_{ G:{\rm maximal \; tree}} \;\;
\sum_{B:{\rm sets \; of \; 8\mbox{-}fold \; bonds}} e^{-S},
\eeq
where
\beqa
S&=&\sum_{(ij)\in G} 12\ln [(x^i-x^j)^2+1]
 +\sum_{(ij)\in B} 6 \ln [(x^i-x^j)^2+1] \n
 &&+\sum_{i<j} \theta(1-(x^{ij})^2)(-4) \ln[(x^{ij})^2(2-(x^{ij})^2))].
\label{eq:tbmodel}
\eeqa
Here we fix the number of 8-fold bonds by hand.
It is enough to check the above mentioned conjecture,
although the number of 8-fold bonds is actually fixed by dynamics.

We generate distribution of $\{ x^i_\mu \}$ by the Monte Carlo method.
We then measure moment of inertia,
\[
M_{\mu\nu} = \sum_{i=1}^{N} (x^i_\mu-x^{\rm cm}_\mu)
                            (x^i_\nu-x^{\rm cm}_\nu),
\]
and diagonalize it.
The ten eigenvalues are the length squared in the principal axes
in ten dimensions.
In this way, we analyze anisotropy of ten-dimensional space-time.
For example, if four of the eigenvalues are much larger than the others,
and grow as we take $N$ large,
it means space-time is four-dimensional.

Figure \ref{fig:eigen} is our preliminary result.
In $N=800$, two of the ten eigenvalues are relatively large,
suggesting the existence of anisotropy.
We hope to see four  eigenvalues become larger as we increase $N$.
However, in order to see this,
we need at least $4^4 \cdot 2^6 \sim 16,000$ points,
which is quite a large number in the current computer power.
Thus, we are trying to make some modifications to the model
(\ref{eq:tbmodel}) to study the system with larger $N$.
The results will be reported soon. \cite{AKN}
\section{Symmetries in the low energy theory}
\setcounter{equation}{0}
\subsection{Local gauge invariance}
Once we describe the  space-time as a
dynamically generated distribution of the eigenvalues,  low-energy effective
theory in the space-time
 can be obtained by solving  the dynamics of the fluctuations around the
background $X_{\mu}$.
Both of  the space-time $X_{\mu}$ and matter $\tilde{A}_{\mu}$
 are unified in the same matrices $A_{\mu}$
and should be determined dynamically.
Low-energy fluctuations
 are  generally composites of $A_{\mu}$ and $\psi$, and
it is natural from the analysis of the
Schwinger-Dyson equation for the  Wilson loops
that a local operator in   space-time
is given by a microscopic limit of the Wilson loop operators,
such as
\beq
w(k;O) ={\rm Tr} [ O(A,\psi ) \exp (i k^{\mu}A_{\mu}) ].
\label{localop}
\eeq
Here, $O(A, \psi )$ is some operator made of $A_{\mu}$ and $\psi$.
In order to identify the total momentum of this operator with
$k$, the operator $O(A, \psi )$ should be invariant under  a constant
shift of $A_{\mu}$, that is, a translation in the
space-time coordinates.
\par
In the first approximation around the diagonal background $X_{\mu}$,
the coordinate representation of this  operator is given by
\beqa
\hat{w}(x;O) &=& \int  {d^{10}k \over (2 \pi)^{10}} 
\exp(- i  k^{\mu} x_{\mu}) \ w(k;O)
\n
&\sim& \sum_{i=1}^{N} O_{ii} \delta^{(10)}({\boldsymbol x}-{\boldsymbol{ x_i}}).
\eeqa
Here, we have replaced $A_{\mu}$ by $X_{\mu}+\tilde{A}_{\mu}$ and taken
the leading term. $O_{ii}$ is the $(ii)$ component of the operator $O$.
Due to the delta function,
the operator has  support   only in the domain where the
eigenvalues distribute.
 Vanishing of the operator $\hat{w}(x;O)$
outside of  the domain of the distributed  eigenvalues
implies that space-time simply does not exist outside
the domain.
This fact supports our interpretation of the space-time
in IIB matrix model.
\par
We can apply a similar analysis to strings which propagate in the space-time.
In the $1/N$ expansion, the
correlation  between Wilson loop operators can be evaluated by
summing over all surfaces made of Feynman diagrams
connecting the Wilson loops at the boundary.
This surface is interpreted as a string world sheet connecting
strings at the boundaries.
Each eigenvalue ($x_{\mu}^i$) associated to
  a loop in  the diagrams represents  a
coordinate on the world sheet,  and
it takes a value in the eigenvalue distribution in the leading
approximation around the diagonal background $X_{\mu}$.
Hence, a string world sheet evolves only in the  space-time of the
eigenvalue distribution, which  again supports our interpretation of
 space-time.
\par
It is generally difficult  to
obtain how fluctuations  propagate in the eigenvalue distribution,
which  is reminiscent of the QCD effective theory:
In QCD excitations are expressed as composite operators of microscopic
variables,
 and their low-energy dynamics can be discussed  only
through  a  symmetry argument,
namely an argument based on chiral symmetry.
Also, in our case,  we will show that there are eigenvalue distributions
around which  symmetry arguments  allow us to discuss the  low-energy
dynamics for some excitations.
Suppose that the
eigenvalue distribution  forms clusters consisting  of $n$ eigenvalues.
At a length scale much larger than the size of each cluster,
the  $SU(N)$ symmetry is
broken down to $SU(n)^{m}$, where $m=N/n$.

We can
expand $A_{\mu}$ and $\psi$ around such a background $X_{\mu}$
similarly to the analysis in the previous section.
First, write $A_{\mu}$ and $\psi$ in  block forms:
\beqa
A_{\mu}&=& \left( \begin{array}{llll}
                 A_{\mu}^{11} &A_{\mu}^{12} & ... &  \\
                 A_{\mu}^{21} &A_{\mu}^{22} & & \\
                  \vdots            &             & \ddots &   \\
                  &&& A_{\mu}^{mm}
         \end{array} \right), \n
\psi &=& \left( \begin{array}{llll}
                 \psi^{11} &\psi^{12}& ...& \\
                 \psi^{21}& \psi^{22}&& \\
                 \vdots & & \ddots &   \\
                  &&& \psi^{mm}
         \end{array} \right).
\eeqa
Each block, $A_{\mu}^{ij}$ or $\psi^{ij}$,
is an $n \times n$ matrix, and  the diagonal blocks
can be further decomposed:
\beqa
A_{\mu}^{ii} &=& x_{\mu}^{i} {\bf 1}+\tilde{A}_{\mu}^{ii}, \n
\psi^{ii} &=& \xi^i {\bf 1}+\tilde{\psi}^{ii},
\eeqa
where $ {\bf 1}$ is an $n \times n$ unit matrix and
tr $\tilde{A}_{\mu}^{ii}=0$. Here, tr means the trace for a
submatrix of $n \times n$.
We interpret each cluster of the eigenvalues
as being a space-time point with an internal
structure $SU(n)$.
Since each $SU(n)$ symmetry acts on the variables at
 position $i$ independently,
the unbroken  $SU(n)^m$ symmetry can be regarded as  being local gauge
symmetry.
Indeed, under a gauge
transformation  of the unbroken $SU(n)^m$ symmetry,
\beqa
g=\left( \begin{array}{llll}
                 g_1 &    &  & \\
                     &g_2 &  & \\
                     &    & \ddots &   \\
                     &    &       & g_m \\
         \end{array} \right)  \in SU(n)^m \subset   SU(N),
\eeqa
the diagonal block fields, $\tilde{A}_{\mu}^{ii}$ and $\tilde{\psi}^{ii}$,
transform as adjoint matters,
(i.e., site variables in  the lattice gauge theory),
while the off-diagonal block fields,   $A_{\mu}^{ij}$ and $\psi^{ij}$,
transform as gauge connections
(i.e., link variables):
\beqa
\tilde{A}_{\mu}^{ii} \rightarrow g_i \tilde{A}_{\mu}^{ii} g_i^{\dagger},
\n
\tilde{\psi}^{ii} \rightarrow g_i \tilde{\psi}^{ii} g_i^{\dagger},
\n
A_{\mu}^{ij} \rightarrow g_i A_{\mu}^{ij} g_j^{\dagger},
\n
\psi^{ij} \rightarrow g_i \psi^{ij} g_j^{\dagger}.
\eeqa

Some of the dynamics for
low-energy excitations is governed by this
local gauge invariance.  Gauge fields live on the links   and transform
as the link variables in  lattice gauge theory. In our case, we have
too many such fields (at least  10 boson fields $A_{\mu}^{ij}$
for a link $(ij)$ ),
but only one unitary link variable
 is  assured to be
massless by the gauge symmetry, and the
others  acquire mass dynamically.
Therefore, in deriving low-energy effective theory,
 we first apply a polar decomposition to $A_{\mu}^{ij}$
into unitary and hermitian degrees of freedom, and
identify all of the  unitary components
of $A_{\mu}^{ij}$
by setting  them to be one common field
 $U^{ij}$ on each link.
We have to integrate those massive off-diagonal block fields
$A_{\mu}^{ij}$ while keeping the unitary components $U^{ij}$
and this procedure is performed by the following replacements:
\beqa
\overline{A_{\mu}^{ij}} &=& 0, \n
\overline{A_{\mu}^{ij} \otimes A_{\nu}^{ji}} &=&
g^2 {\delta_{\mu \nu} \over (x^{ij})^2} U^{ij} \otimes U^{ji}.
\label{replacement}
\eeqa
In the second equation, the  factor, $1/(x^{ij})^2$,
corresponds to the
propagation of the hermitian degrees of freedom while the
appearance of the link variable $U^{ij}$ corresponds to keeping the unitary
degrees of freedom.
Higher order correlations can be obtained by using Wick theorem
in general except ten-body correlation function for $A_{\mu}^{ij}$.
Due to the chiralness of the ten-dimensional fermion, we obtain
an extra term proportional to $\epsilon_{\mu_1 ... \mu_{10}}$,
coming from fermion one-loop integral.
\par
In order to derive the effective theory for fluctuations
around the assumed background, we integrate massive fields
first and obtain effective action for other fields
as we have obtained the effective action for the diagonal
components (i.e., space-time coordinates) in the previous section.
Generally speaking, we can expect any terms which are not
forbidden by symmetries; supersymmetry and local $SU(n)$ gauge
symmetry.
\par
A plaquette action for gauge fields $U^{ij}$ can be generated as follows.
The relevant terms in the action are
\beq
S_{4}= {1 \over g^2} \sum_{i \ne j \ne k \ne l} {\rm tr}
(A_{\mu}^{ij} A_{\nu}^{jk} -A_{\nu}^{ij} A_{\mu}^{jk})
A_{\mu}^{kl} A_{\nu}^{li}.
\eeq
By integrating out the hermitian degrees of freedom of the
off-diagonal blocks
with the procedure  (\ref{replacement}), this action vanishes;
$\bar{S_4}=0$. However,  interactions
generated by $(S_4)^2$ induce a  kinetic term for the gauge
field:
\beq
\overline{(S_4)^2} \sim  \sum_{i \ne j \ne k \ne l}
{g^4 \over (x^{ij})^2(x^{jk})^2 (x^{kl})^2 (x^{li})^2  }
{\rm tr}(U^{ij} U^{jk} U^{kl} U^{li}) {\rm tr}(U^{il}U^{lk}U^{kj}U^{ji}).
\label{plaquette}
\eeq
This is the  plaquette action generated by a Wilson loop for
the adjoint representation, and hence the  gauge field $U^{ij}$
indeed propagates in  the space-time of the eigenvalue distribution.
The gauge field can hop between any pair of space-time
points, but the hopping is suppressed by $1/x^8$ for distant points
and we will recover locality in the continuum limit.
Similarly we can obtain a gauge invariant hopping term
for adjoint fermion $\psi^{ii}$.
\par
To summarize this subsection,
supposing  that  the distribution of the eigenvalues  consists of
small clusters of size $n$, we have shown that the low-energy
effective theory contains several massless fields, such as
the gauge field associated with the local $SU(n)$ gauge symmetry
and  fermion  field in the adjoint representation of $SU(n)$
gauge symmetry. Gauge-invariant kinetic terms were also
derived.
Our findings are reminiscent of those which are obtained by
considering $n$ coincident D9-branes. Presumably our argument here
is related to the standard argument of coincident D-branes.
\par
Our system  is a lattice gauge theory
on a dynamically generated random lattice.
It is invariant under a permutation for the set of the $m$ discrete
space-time points,
since the permutation group $S_{m}$ is a subgroup of
the original $SU(N)$ symmetry.
The existence of the permutation symmetry
is the crucial difference from the ordinary lattice gauge theory
on a fixed lattice,
which   becomes important in deriving the
diffeomorphism invariance of our model.
We will come back to this point in the next subsection.
Although the permutation invariance requires
that all space-time points are equivalent,
 locality in  the space-time will
be assured due to suppression of the hopping term between  distant
points. In general, however, we need  a sufficient power for  the damping
of the  hopping terms in order to assure locality in the continuum limit.
Though  we do not yet know the   real condition for locality,
we expect that  terms with lower  powers
are canceled due to   supersymmetry or
by averaging over gauge fields.
\par
\subsection{Diffeomorphism invariance}
As shown in \S2, the one-loop effective action for
the space-time points  is described as a statistical
system of $N$  points whose coordinates are $x_{\mu}^{i}$.
Integration over the fermion zeromodes $\xi$ gives
the Boltzmann weight, which depends on a graph (or network) connecting
the space-time  points locally by  the bond interactions:
 \begin{equation}
Z= \sum_{G:{\rm graph}} \int dX \ W[X;G].
 \end{equation}
$ W[X;G]$ is a complicated function of  a configuration $X$ and a
graph $G$. An important property is that the weight is suppressed
at least by a damping factor of $1/(x^i-x^j)^{12}$,
when two  points, $i$  and $j$,
are connected.
This system is, of course,  invariant under permutations $S_N$\footnote{In this section we consider general eigenvalue distributions
in which all eigenvalues have nondegenerate  space-time coordinates.
If we take the cluster type distribution considered in the previous
subsection, the permutation symmetry responsible for the diffeomorphism
invariance should be $S_{N/n}.$
}
of $N$ space-time
points, which is a subgroup of the original symmetry $SU(N)$,
while  the  Boltzmann weight for  each graph $G$ is not.
The invariance  is realized by summing  over all possible graphs.
In other words, the system becomes permutation invariant by
rearrangements of the bonds in the  network of the space-time points.
This reminds us of  the dynamical triangulation approach to
quantum gravity, \cite{ambjorn}
where  diffeomorphism invariance is believed to arise from
summing all possible triangulations.
Our system satisfies both the locality
and permutation invariance simultaneously
by summing  over  all possible graphs whose points are
connected through the local interactions.
\par
Now  we  see that the permutation
invariance of our system actually leads to  diffeomorphism invariance.
To see how  the background metric  is encoded
in the  effective action for low-energy excitations,
let us consider, as an example,  a  scalar field $\phi^i$ propagating in
distributed eigenvalues.  The effective action will be given by
\begin{equation}
S= \sum_{i,j} {(\phi^i-\phi^j)^2 \over 2} f(x^i-x^j) + \sum_{i} m (\phi^i)^2,
\end{equation}
where $f(x)$ is a function  decreasing  sufficiently fast at infinity
to assure
locality in  the space-time. Introducing  the density function of the
eigenvalues,
\begin{equation}
\rho(x)= \sum_i \delta^{(10)}(x-x^i),
\end{equation}
and a  field $\phi(x)$ which  satisfies
 $\phi(x^i)=\phi^i$,
the action can be rewritten as
\begin{equation}
S = \int dx dy \langle \rho(x) \rho(y) \rangle
{(\phi(x)-\phi(y))^2 \over 2} f(x-y)
+ m \int dx  \langle \rho(x) \rangle \phi(x)^2.
\end{equation}
Here,  the expectation, $\langle \cdots \rangle$,
for the density and the density  correlation  means that we have
taken average over  configurations $X$ and   networks  $G$ of
the space-time points.

Normalizing  the density correlation in terms of the density,
\begin{equation}
\langle \rho(x) \rho(y) \rangle =
\langle \rho(x) \rangle \langle \rho(y) \rangle
r(x,y),
\end{equation}
and expanding $\phi(x)-\phi(y) = (x-y)_{\mu}\partial^{\mu} \phi(x) +
\cdot \cdot \cdot$, the action becomes
\begin{eqnarray}
S &=& {1 \over 2} \int dx \langle \rho(x) \rangle  \left[  \int dy
\langle \rho(y) \rangle (x-y)_{\mu}(x-y)_{\nu} f(x-y) r(x,y) \right] 
\partial^{\mu} \phi(x) \partial^{\nu} \phi(x)\nonumber\\
&& + m \int dx \langle \rho(x) \rangle \phi(x)^2 \cdots .
\label{phiaction}
\end{eqnarray}
This expansion shows  that  the field $\phi(x)$ propagating
in  the eigenvalue distribution
feels  the density correlation as the  background metric,
while  the density itself as  vacuum expectation value of
the dilaton field. Namely, we can identify
\begin{eqnarray}
g_{\mu \nu}(x) &\sim&  \int dy
\langle \rho(y) \rangle (x-y)_{\mu}(x-y)_{\nu} f(x-y) r(x,y),
\label{metric} \\
\sqrt{g} e^{- \Phi(x)} &\sim& \langle \rho(x) \rangle.
\label{dilaton}
\end{eqnarray}
If the density correlation respects the original translational and
rotational symmetry, that is, if they are not spontaneously broken,
the metric becomes flat, $g_{\mu \nu} \sim \eta_{\mu \nu}$.
(Normalization can be absorbed by the dilaton vev.)
The fact that the background metric is  encoded in the density
correlations  indicates that our system is general covariant,
even though the IIB matrix model action (\ref{action}) defined
in flat ten dimensions does not have  a manifest general covariance.
\par
Then, let us see how the diffeomorphism invariance  is realized in our
model.
The action  (\ref{action})
is invariant under the permutation $S_N$ of the eigenvalues, which is a
subgroup of $SU(N)$.
Under a permutation,
\begin{equation}
x^i \rightarrow x^{\sigma(i)} \ \ \mbox{for} \ \ \sigma \in S_N,
\label{perm}
\end{equation}
the field $\phi^i$ transforms into $\phi^{\sigma(i)}$.
Then, from the definition of the  field $\phi(x)$, we should
extend the transformation (\ref{perm}) into $x$,
\begin{equation}
x \rightarrow \xi(x),
\label{permx}
\end{equation}
such that $\xi(x^i) = x^{\sigma(i)}$.
Under this transformation, the eigenvalue density transforms as
a scalar density and the  field $\phi(x)$ as a scalar field.
On the other hand, the metric transforms as a second-rank tensor, if the
function $f(x-y)$ decreases rapidly around $x=y$
and the $y$ integral in Eq.~(\ref{metric})
has support  only near $y=x$.
The tensor property of the metric
 is also required  from the invariance of the
action under  transformation (\ref{permx}).
In this way,  the
invariance under a permutation of the eigenvalues leads to
the invariance of  the low-energy effective action
under general coordinate transformations.
\par
The background metric is encoded in the  density correlation of the
eigenvalues. Since we have started  from the
IIB matrix model action (\ref{action}) which is Poincar\'e-invariant,
the density correlation is
expected to be translational and rotational invariant,
and we may  obtain a low-energy effective action  in a flat background.
A nontrivial background can be induced dynamically
if the Lorentz symmetry is spontaneously broken and
the  eigenvalues are nontrivially distributed.
\par
A nontrivial background can  also be described  by condensing a graviton
opera-\linebreak
tor. \cite{yoneya-nishi}
Bosonic parts of graviton and dilaton operators are given by
\begin{eqnarray}
S_{\mu \nu}(k) &\sim&   {\rm Tr}(F_{\mu \lambda} {F^{\lambda}}_{\nu} e^{ik \cdot A})
+ (\mu \leftrightarrow \nu),  \label{graviton}
\\
D(k) &\sim&   {\rm Tr}(F^2 e^{ik \cdot A}). \label{dilaton2}
\end{eqnarray}
Their condensation  induces extra terms in the IIB matrix model action,
\begin{equation}
S_{\rm{cond}} = \int dk\left (\sum h^{\mu \nu}(k) S_{\mu \nu}(k) + h(k) D(k)\right).
\end{equation}
We can similarly obtain an effective action for fluctuations  around a diagonal
background from this modified matrix model action.
 Condensation of dilaton changes
the Yang-Mills coupling constant $g$ locally in  space-time.
Since $g$ is the only dimensionful constant in our model, and thus determines
the fundamental length scale, a local change in $g$  leads to
a local change in the eigenvalue density.
This is consistent with our earlier discussion that  the dilaton
expectation value is encoded in the  eigenvalue density.
On the other hand, the condensation of graviton induces an asymmetry of
space-time. For a condensation of the $k=0$ graviton mode, it is
obvious that the condensation can be compensated by a field
redefinition of matrices $A_{\mu}$,
\beq
A_{\mu} \rightarrow (\delta_{\mu}^{ \nu} + h_{\mu}^{ \nu}) A_{\nu},
\eeq
and the two models, the original IIB matrix model and the modified
one with the $k=0$ graviton condensation,
are directly related through the above field redefinition.
The density of the eigenvalues is mapped accordingly, and the density
correlation is expected to become asymmetric in the
modified matrix model.
For a more general condensation,
if  the  graviton operator $\hat{S}_{\mu\nu}(x)$
(coordinate representation of  Eq.~(\ref{graviton}))
 changes only  the local property of the dynamics of the eigenvalues,
 the  density correlation will become asymmetric locally in space-time
around  $x$,
and therefore induces a local change in the background metric.
\par
Our low-energy effective action is formulated as a lattice gauge theory
on a dynamically generated random lattice. Since  the lattice itself is
generated dynamically from  matrices, we must sum over all possible
graphs.  In this way,  our system
is permutation $S_N$ invariant,  which is responsible for the
diffeomorphism invariance.
The background metric is encoded in the density correlation of the
eigenvalues, and the low-energy effective action becomes manifestly
general covariant.
The graviton operator is represented as fluctuation around the
background space-time, and is
constructed from the off-diagonal components of the
matrices. A microscopic derivation of the
propagation of the graviton is difficult to obtain,
but  once we have clarified the
underlying diffeomorphism symmetry,
it is natural that the low-energy effective action for the graviton
is described by the Einstein Hilbert action.
By employing this diffeomorphism invariance and the supersymmetry,
we will be able to  derive the low-energy behavior of the graviton multiplets,
which will be reported in a separate paper.
\section{Discussion}
We have reviewed the current status of the type IIB matrix model,
which is proposed as a constructive definition of superstring.
\par
There are still  several  conceptual issues.
We have obtained a nonabelian gauge symmetry from the IIB matrix model
by assuming a particular eigenvalue distribution.
This indicates that this  vacuum  is not a perturbative vacuum
of the type IIB superstring. Instead, we may wonder if this is
 a perturbative vacuum of a heterotic string or a type I string realized
in a nonperturbative way within the IIB matrix model.
As we discussed in the introduction,  our matrix model
contains both of the world sheets of the
fundamental  IIB string and those of the D-strings.
By a semiclassical correspondence (\ref{correspondence}),
we have identified a
 IIB superstring in the Schild gauge where
 tr is interpreted  as  integration over a D-string world sheet.
We can also construct an  F-string world sheet
in terms of surfaces made of
Feynman diagrams, whose $SU(N)$ index  represents
the space-time coordinate of a world sheet point.
In both cases, if we assume an eigenvalue distribution consisting of
small clusters, an internal structure appears on the world sheet,
and hence current algebra may arise and there is a possibility to
describe heterotic string within the type IIB matrix model.
\par
Another issue is how to describe global topology in the IIB matrix
model. A simple example is a torus compactification.
A  possible procedure of  torus compactification \cite{Taylor}
is  to identify $A_{\mu}$ with $A_{\mu} + R_{\mu}$
by embedding a derivative operator into our matrix configuration.
Therefore,  $N$ is taken as infinity from the beginning.
Since this procedure has a subtlety in the large $N$ limit,
we require a careful examination of the double scaling limit.
\par
We also do not yet know how we can describe chiral fermions
in lower dimensions after compactification.
If we naively consider a low energy effective theory on a 
four-dimensional space-time generated by distributed eigenvalues,
we will obtain fermions with both chiralities.
A possible mechanism to produce a chiral fermion is to consider
a compactified six-dimensional space-time with a nontrivial index,
or a parity violating background.
We should  replace the $n \times n$
block matrices considered in \S4 by $(n+\infty) \times (n+\infty)$
matrices. Size $n$ part of a block represents the low energy gauge symmetry
as before. The rest of the infinite size  represents an internal space
with six dimensions, which should have a non-trivial index.
For simplicity, let us consider a two-dimensional internal space.
Then the following background
\beqa
A_{a}^{ii} &=& x_{a}^{i} {\bf 1},\n
A_{5}^{ii} &=& \left( \begin{array}{ll}
                0 & \n
                  & P
         \end{array}
                         \right), \n
A_{6}^{ii} &=& \left( \begin{array}{ll}
                0 & \n
                  & Q
         \end{array}
                         \right)
\eeqa
gives the $(ii)$ block components of a desired background.
Here $a=1 \sim 4$ and $P, Q$ are infinite dimensional matrices
satisfying $[P, Q] = -i$.
This background is invariant under $U(n) \times U(1)$ in each
block and this becomes the local gauge symmetry in four-dimensional
space-time.
The effective theory for low energy fluctuations
 around the background are similarly  obtained
as in \S 4.
All the off-diagonal fields become massive and we integrate them over
except a gauge field.
In diagonal blocks, there are several fields that can be massless.
Writing  fluctuations  in a diagonal block, $O =  \tilde{A}_{\mu}^{ii}$ or
$\tilde{\psi}^{ii}$ as
\beqa
O = \left( \begin{array}{ll}
        O_{A}    &  O_{F} ^{\dagger}\\
        O_{F}  &  O_{S}
         \end{array}
                         \right), 
\eeqa
$O_{A}$ transforms as an adjoint representation,
$O_{F}^{\dagger}$ as a fundamental representation and $O_{S}$  is a singlet
for $SU(n)$ transformation.
$P$ or $Q$ acts  like  $P O_{F}^{\dagger}$ on $O_{F}^{\dagger}$,
and $[P, O_{S}]$ on  $S$-components.
Then we can show that we obtain a massless chiral fermion with
fundamental representation for gauge symmetry $SU(n)$,
whose wavefunction in $(P,Q)$ space is given by the groundstate
wavefunction of a harmonic oscillator.
Other possibly massless fermions are vector-like and will acquire
mass unless they are protected by supersymmetry.
This is the simplest way to obtain a chiral fermion in four-dimensional
space-time. In this construction, all non-singlet fields transforming
as adjoint or fundamental representations live in four dimensions
and localized at $x_5=x_6=0$. However, singlet fields including graviton
propagate in the bulk (here, six dimensions).
In order to have a  four-dimensional theory, we need to
compactify the internal six-dimensional space.
This is discussed in a separate paper.
\par
It is also desirable to construct $AdS$ type backgrounds in our
approach.\cite{AdS}
Let us recall the metric of $AdS_5 \times S^5$:
\beq
ds^2 = {R^2\over z^2}(d\vec{x}^2+dz^2)+R^2d\Omega^2_5 .
\eeq
The volume factor $\sqrt{g} \sim {1\over z^5}$ is sharply
peaked at $z=0$ in $AdS_5$. Since we have argued that $\sqrt{g}$ is
proportional to the density distribution in IIB matrix model, such a
background may be represented by the eigenvalue distribution which
is also sharply peaked at $z=0$ namely at the four-dimensional boundary.
The gauge theory is obtained by assuming that the full $SU(N)$ matrices are
decomposed
into the clusters of submatrices of $SU(n)$ as we have argued.
Since the eigenvalue distribution is essentially four-dimensional,
the resulting low energy effective theory must be a four-dimensional field
theory.
If we further assume ${\cal N}=4$ supersymmetry in four dimensions which
must be present due to the conformal symmetry of $AdS_5 \times S^5$,
we may conclude that the low energy effective theory for such a background
of IIB matrix model is ${\cal N}=4$ super Yang-Mills theory.


\end{document}